# Sodium induced beneficial effects in wide bandgap Cu(In,Ga)S$_2$ solar cell with 15.7% efficiency

Arivazhagan Valluvar Oli[1], Kulwinder Kaur[1] Michele Melchiorre[1], Aubin Jean-Claude Mireille Prot[1], Sevan Gharabeiki[1], Yucheng Hu[2], Gunnar Kusch[2], Adam Hultqvist[3], Tobias Törndahl[3], Wolfram Hempel[4], Wolfram Witte[4], Rachel A. Oliver[2] and Susanne Siebentritt[1]

[1]Laboratory for Photovoltaics, Department of Physics and Materials Science Research Unit, University of Luxembourg, 41 rue du Brill, L-4422 Belvaux, Luxembourg
[2]Department of Materials Science and Metallurgy, University of Cambridge, 27 Charles Babbage Road, Cambridge CB3 0FS, UK
[3]Ångström Solar Center, Division of Solar Cell Technology, Department of Materials Science and Engineering, Uppsala University, 75121 Uppsala, Sweden.
[4]Zentrum für Sonnenenergie- und Wasserstoff-Forschung Baden-Württemberg (ZSW), Meitnerstraße 1, 70563 Stuttgart, Germany

Email: arivazhagan.valluvaroli@gmail.com ; susanne.siebentritt@uni.lu

**Abstract**

This study underscores the pivotal role of sodium (Na) supply in optimizing the optoelectronic properties of wide bandgap (~1.6 eV) Cu(In,Ga)S$_2$ (CIGS) thin film absorbers for high efficiency solar cells. Our findings demonstrate that the synergistic use of Na from the glass substrate, in conjunction with in-*situ* sodium fluoride (NaF) co-evaporation, significantly enhances the structural and optoelectronic properties of the CIGS. CIGS grown under either Na-deficient or excess conditions exhibits inferior microstructural and optoelectronic properties, whereas an optimal Na supply leads to enhanced photovoltaic performance. Optimal Na incorporation minimizes vertical gallium fluctuations and improves the grain size and crystallinity. An absolute 1 sun calibrated photoluminescence (PL) measurement reveals a substantial suppression of bulk defects and a reduction in non-radiative losses, resulting in a high quasi-fermi level splitting ($\Delta E_F$) of 1.07 eV, 93 meV higher than in Na-deficient CIGS with the same bandgap. Optimal Na supply further increases excited carrier decay time, as revealed from time-resolved PL, and hole doping density. Cross-sectional hyperspectral cathodoluminescence mapping reveals that optimal Na supply significantly reduces defect density near the surface, thereby effectively translating $\Delta E_F$ to open-circuit voltage ($V_{OC}$). As a result, a champion wide bandgap CIGS solar cell with a cadmium-free ZnSnO buffer layer achieved an impressive $V_{OC}$ of 971 meV and an active area power conversion efficiency of 15.7%, highlighting its potential for advancing tandem photovoltaic technologies with stable inorganic top cell.

**Introduction**

Sulfide-based chalcopyrite with the composition Cu(In,Ga)S$_2$ (CIGS) is particularly promising as a top cell in tandem solar cell configurations due to its tunable direct bandgap (E$_g$), ranging from 1.55 eV (CuInS$_2$) to 2.4 eV (CuGaS$_2$) [1]. Chalcopyrite solar modules have demonstrated long-term environmental stability in field tests [2, 3], making them a viable candidate for commercialization. Additionally, the growth techniques used for CIGS, such as physical vapor deposition, can be seamlessly integrated into existing large-scale manufacturing processes employed for low-bandgap solar cells based on materials like Cu(In,Ga)Se$_2$ (CIGSe) [4, 5]. Despite the progress in increasing the power conversion efficiency (PCE) of CIGS to over 15% [6-8], it still lags behind its selenium (Se) counterpart, which has recently achieved a certified PCE of 23.6% [9]. The primary factor limiting the performance of CIGS solar cells is the open-circuit voltage (V$_{OC}$) deficit [10, 11]. The V$_{OC}$ is closely related to the quasi-Fermi level splitting ($\Delta E_F$) of the photovoltaic absorber, which represents the maximum V$_{OC}$ that a solar cell can achieve with ideal interfaces [12, 13]. Non-radiative recombination channels, which arise from defects within the absorber and at the interfaces, contribute to a $\Delta E_F$ deficit (referred to as 'non-radiative loss' throughout the text) defined as the difference between the ideal Shockley-Queisser (SQ) V$_{OC}$ and $\Delta E_F$ [7, 14, 15]. This non-radiative loss negatively impacts the minority carrier lifetime ($\tau$) and thus V$_{OC}$ in Cu(In,Ga)S$_2$ solar cells [7, 16, 17]. Therefore, understanding and mitigating these defects is crucial for obtaining high-quality absorbers [12, 13]. Interface recombination is another important factor which limits the V$_{OC}$ due to band misalignment and interface defects [18, 19]. Strategies such as substrate temperature control, composition engineering, and doping have been widely employed to achieve high-quality sulfide solar cell absorbers [6, 7, 20-22]. CIGS grown at high deposition temperatures and with copper (Cu)-poor stoichiometry has achieved the highest reported PCE of over 15% [6-8]. High deposition temperature increases the grain size due to improved elemental diffusion, while Cu-poor surface compositions helps suppress the deep defects and improve the solar cell performance [7, 20].

Sodium (Na) has been known to play a vital role on the improved performance of selenide-based chalcopyrite solar cells through various mechanisms [23-27]. In CIGSe, Na is typically introduced through diffusion from soda-lime glass (SLG) substrates, precursor layers, post-deposition treatments (PDT), or co-evaporation during growth [24, 26, 28, 29]. The presence of Na improves CIGSe absorber performance by mechanisms including substitution of antisite donor defects (e.g., $Na_{In}$ or $Na_{Cu}$), formation of Cu vacancy ($V_{Cu}$), passivation of grain boundaries, enhanced atomic diffusion, increased p-type conductivity, and grain size enlargement [30-34]. However, excessive Na content in the absorber can be detrimental to solar cell performance [35-37]. The method of Na incorporation also significantly influences its effect. For example, a statistical comparison in CIGSe found that NaF PDT yielded better absorber quality and solar cell performance compared to the same amount of NaF co-evaporated during growth [27], however, to our knowledge, no similar comparative study has so far been published that addresses the impact of Na in CIGS. A theoretical study on Na's effect within CIGSe grains proposes that Na supplied via SLG or PDT forms $Na_{Cu}$ antisites during high-temperature growth, which then diffuse to grain boundaries (GB) upon cooling, creating Cu vacancies that increase p-type conductivity [32]. Observation of indium (In) and gallium (Ga) inter grain diffusion promoted by Na revealed that Na segregation at CIGSe GB plays a decisive role by acting as a barrier for In/Ga inter grain diffusion in CIGSe absorber [33]. Given the similarities between sulfide and selenide chalcopyrites, the effects of Na in CIGSe provide valuable insights for understanding its role in CIGS [10]. Similar to CIGSe, Na diffusion from SLG during CIGS growth has been found to enhance hole doping density and yield higher PCE than low-Na absorbers [38]. Additionally, Cu-poor CIGS absorbers grown with optimal NaF precursor layers exhibited improved $V_{OC}$, which decreased with further increases in NaF thickness [22]. Preliminary investigations suggest that Na diffusion from SLG and NaF co-evaporation during CIGS growth positively affect $\Delta E_F$, while the precursor layer approach had a negative impact [10]. Despite these insights, the specific effects and optimal methods of Na incorporation in CIGS remain underexplored, highlighting the need for further comparative studies in this area.

Here, we study the impact of Na addition on the optoelectronic and structural properties of CIGS, particularly its influence on defects, recombination dynamics, non-radiative losses, $\Delta E_F$, and grain size. Specifically, radiative recombination and defect concentration across the depth of graded bandgap CIGS absorbers have not yet been investigated. This study addresses these gaps by systematically introducing Na into the absorber and examining its effects from optoelectronic, microstructural, and solar cell perspectives. We investigated three different methods to control the Na supply: using a low Na containing glass (LNG) substrate, using an SLG substrate, and adding extra Na via NaF co-evaporation during the second stage of CIGS growth on SLG. We explored Na's impact on the graded bandgap absorber through elemental depth profiling and correlated these findings with microstructural and optoelectronic characterizations. We employed absolute calibrated photoluminescence (PL) measurements, time-resolved PL, and cathodoluminescence (CL) spectroscopy to study optoelectronic transitions and address various defects and non-radiative losses. Furthermore, hyperspectral CL cross-sectional mapping was extensively used to understand recombination paths across the absorber depth and investigate the defect concentration near the space-charge region. Our results demonstrate that unintentional Na supply from the substrate during CIGS growth is crucial for reducing deep defects. However, introducing Na intentionally and in a controlled manner through co-evaporation during growth provides additional benefits by reducing bulk defects, minimizing non-radiative losses, and increasing grain size and hole doping density. Furthermore, by identifying and mitigating defect density near the surface, we enhance the translation of the absorber's $\Delta E_F$ into higher $V_{OC}$ in the final solar cell, with reduced interface losses. Finally, we report that a champion solar cell fabricated with a cadmium-free buffer layer achieved a $V_{OC}$ of 971 meV and an active area PCE of 15.7%. This work paves the way to mitigating the $V_{OC}$ deficit, a key challenge in CIGS, and positions CIGS as a promising wide-bandgap absorber for the development of high efficiency tandem photovoltaics.

**Results and discussion**

The CIGS absorbers were deposited using a three-stage evaporation process (refer to the deposition profile in **Fig. S1**). To explore the effects of different Na supplies, we utilized two distinct substrates: LNG with low Na content and SLG with relatively high Na content. Na from these substrates diffuses out at high temperatures, serving as a source during CIGS growth [24]. To introduce additional Na, further samples were prepared by co-evaporating NaF during the second stage of CIGS growth on SLG, with source temperatures varying from 600 °C to 750 °C in 50 °C intervals [39]. Increasing the NaF source temperature, while keeping the evaporation time constant, results in a higher Na supply in the absorber during growth. Above an NaF source temperature of 750 °C, the absorbers began to peel off upon removal from the vacuum chamber, an issue also reported in CIGSe absorbers prepared via the precursor method when exposed to excess NaF [40]. We assessed the quality of all the absorbers using absolute PL measurements (Fig. S7) and found that the 650 °C NaF condition resulted in minimized non-radiative recombination losses, supporting its identification as the optimal Na supply. For simplicity and to facilitate better comparisons, we present the results as LNG (Na-poor), SLG (standard Na), 650

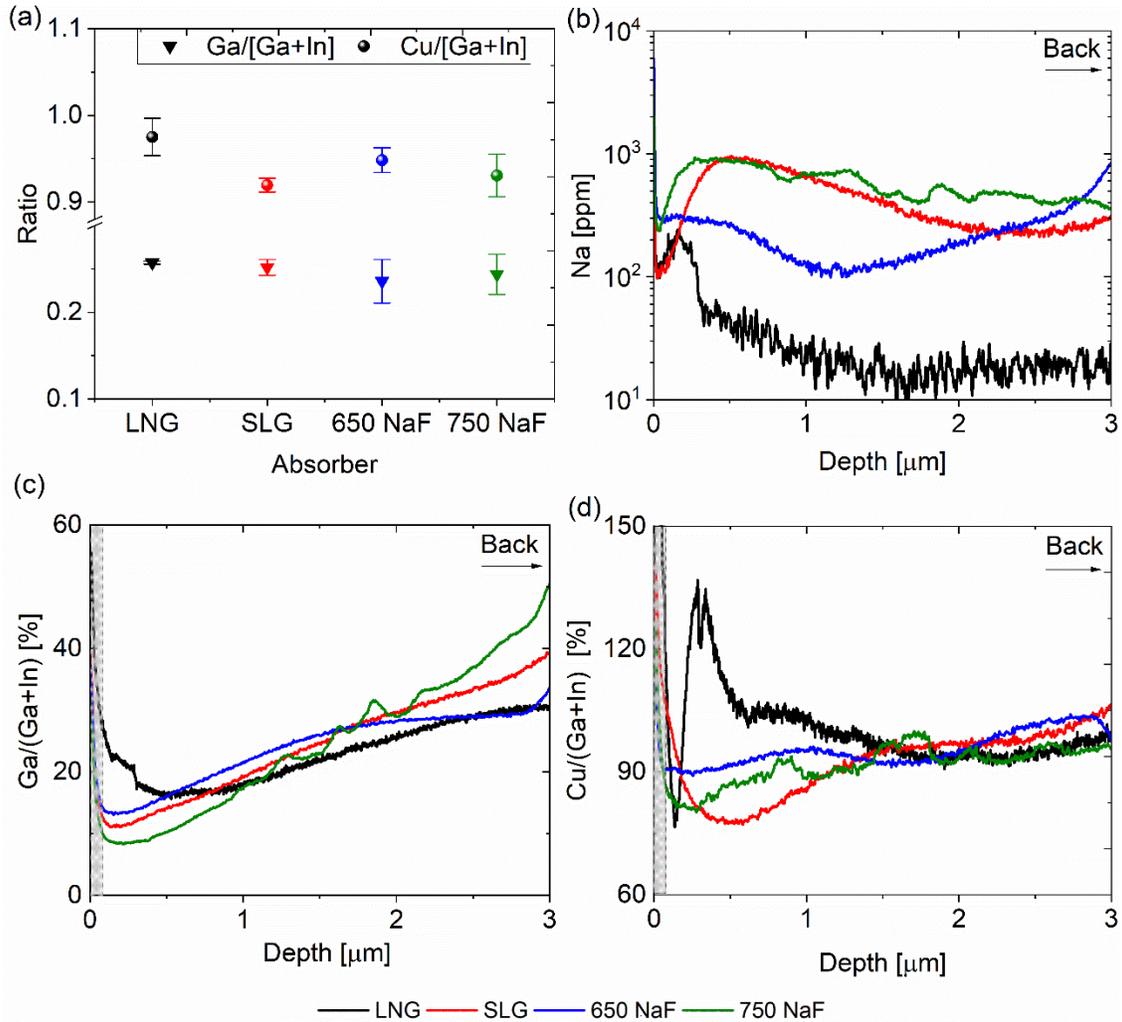

**Fig. 1:** *Elemental composition of CIGS with varying Na supply during growth. (a) Bulk [Ga]/([Ga+In]) and [Cu]/([Ga+In]) ratios obtained from EDS. GDOES depth profile of (b) Na concentration, (c) GGI ratio and (d) CGI ratio. The data points in the near surface region highlighted with a grey rectangle in (c) and (d) are mainly due to artifacts.*

NaF (optimal Na), and 750 NaF (excess Na) throughout the main text. Some results for 600 °C and 700 °C NaF are included in supplementary information in **Fig. S1-S12** and will be referred to in the main text as appropriate.

Energy dispersive X-ray spectroscopy (EDS) was conducted at 20 kV to estimate the bulk composition of the absorbers. **Fig. 1a** presents the bulk composition in terms of [Cu]/([Ga]+[In]) (CGI), [Ga]/([Ga]+[In]) (GGI) ratios. The surface CGI and GGI ratios were measured at 7 kV to capture the signal from the topmost absorber layer, are shown in **Fig. S2**. No significant differences between near surface and bulk composition are found. The compositions of all absorbers are within a similar range, except for LNG, which exhibits a slightly higher CGI. A CGI < 1 indicates a Cu-poor stoichiometry, necessary for optimal solar cell operation [7, 22]. Glow discharge optical emission spectroscopy (GDOES) was employed to trace the elemental depth profile of the absorbers. The absorber depth was estimated using the onset Mo signal from GDOES (**Fig. S3**) and correlated with thickness measurements obtained from cross-sectional SEM images (**Fig. 2**). **Fig. 1b** displays the vertical Na concentration profiles (in ppm) of the absorbers. LNG exhibits negligible Na concentration, while SLG shows a high concentration, confirming that Na diffused from the substrate. This observation is consistent with previous studies on CIGSe absorbers grown on alkali-free and SLG substrates, where Na presence was traced through binding energy or depth profiling [25, 27, 28, 40]. The 650 NaF absorber shows slightly lower Na levels than SLG, whereas the 750 NaF absorber exhibits higher Na concentration throughout the absorber depth. We later discuss a possible explanation for the lower Na concentration in the 650 NaF absorber, relating it to increased grain size and lower density of grain boundaries.

**Fig. 1c** and **1d** show the GGI and CGI depth profiles of the absorbers, respectively. The high signal at the surface (indicated by the grey box) is associated with artifacts, complicating the determination of GGI and CGI variations near the surface. The LNG absorber shows a broad notch, a minimum GGI point which represents $E_g$ in a graded absorber, located around 0.5 µm from the surface, while the SLG absorber displays a sharp notch close to the surface, with steep grading toward the back contact. This grading toward the back contact is intentional in three-stage-grown chalcopyrite absorbers, since conduction band edge grading can prevent photogenerated electrons from reaching the back contact [41]. The magnitude of this grading primarily depends on factors such as gallium flux over time, substrate temperature, and dopants like Na or silver as reported for selenide chalcopyrite absorbers [33, 36, 42]. Adding 650 NaF does not affect the notch position but does shift the GGI minimum upward and reduces the vertical GGI grading. The GGI grading gradually increases from the notch through more than half of the absorber depth, then flattening slightly toward the back contact. However, with 750 NaF, the notch widens, shifts downward, and the GGI grading becomes steeper toward the back contact. This result is consistent with literature comparing Na-free and Na-containing CIGSe absorbers, where variations in Ga concentration were evident in the absorber depth for samples with high Na content [29, 33, 43]. One possible reason for the Na-modulated GGI depth profile variation is improved intragrain diffusion of Ga/In, as Na impedes Ga/In intergrain diffusion by segregating at grain boundaries [33]. We speculate that an abrupt increase in CGI near the surface of the LNG absorber in **Fig. 1d** leads to $CuS_x$ segregation due to insufficient Ga/In intragrain diffusion, resulting from limited Na supply during growth. **Fig. S3** presents the S, Cu, In, and Ga elemental depth profiles for all absorbers. It is evident that all the absorbers are In-rich near the front surface, except for LNG, and Ga rich near the Mo back contact, in agreement with the three-stage deposition profile. Sulfur (S) and Cu are almost uniformly distributed across the depth relative to Ga and In. The fluctuations observed in the 750 NaF absorbers is due instability of plasma during the measurement. For double-graded absorbers, an increase in GGI near the surface, as seen in LNG, is preferred. However, all other absorbers exhibit low GGI grading with a less pronounced notch, which would require nearly perfect passivation of the interface with the buffer layer for optimized solar cell operation.

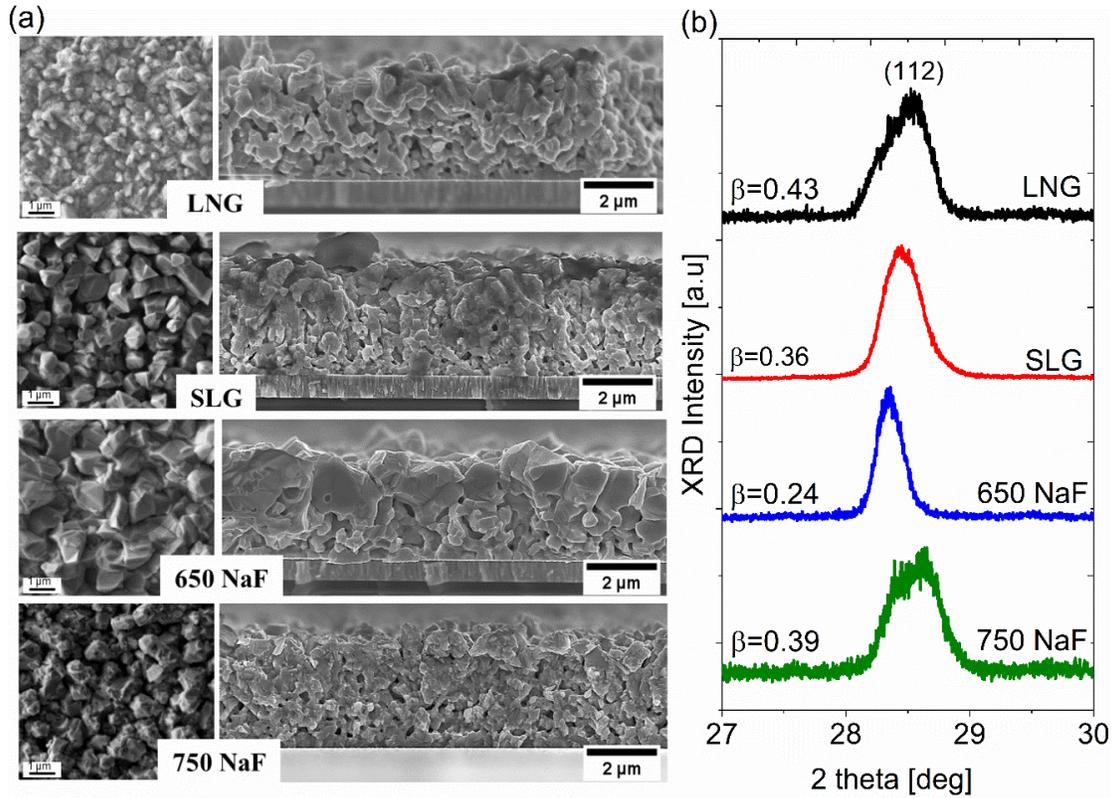

**Fig. 2:** *Microstructure of CIGS with different Na supplied during growth. (a) SE images showing top (left) and cross-sectional (right) views. The scale bar in top view images is 1 µm. (b) X-ray diffractograms of the 112 reflex with corresponding full-width half maximum (β) values.*

**Fig. 2a** displays top view (left) and cross-sectional (right) scanning electron microscopy (SEM) images of CIGS absorbers with different Na supplies. The absorber grown on LNG exhibits smaller surface grains, whereas the grain size increases with the use of SLG substrate, consistent with literature based on selenides [24]. However, in the cross-section, the SLG appears to have grain structures similar to those of the LNG, which may be attributed to inhomogeneities within the SLG absorber. The addition of NaF at a temperature of 650 °C enlarges the grain size, so that grains almost extendfrom top to bottom in a columnar structure as observed in the cross-sectional images. Conversely, 750 NaF reduces the grain size, leading to smaller grains and exhibiting a rough morphology (see enlarged SE image in **Fig. S4**). Cross-sectional SEM images for 600 and 700 NaF are shown in **Fig. S4**, which resemble the SLG and 750 NaF morphological features, respectively. Generally, moderate Na supply during CIGSe growth has been found to increase grain size. For example, the timing of NaF co-evaporation during a three-stage CIGSe growth has been shown to affect microstructure [29]. A comparative study of CIGSe absorbers grown on SLG with and without an alkali barrier layer blocking Na diffusion from the substrate during growth revealed that those with the barrier had smaller grains [39]. Additionally, CIGSe grown on fused silica substrates with extra Na exhibited increased grain size and columnar structure, highlighting the importance of Na supply during growth [28]. Previous studies have indicated that NaF PDT results in larger CIGSe grains compared to NaF co-evaporation, which produces smaller grains in the bulk due to higher Na concentration [36]. However, the exact influence of Na on CIGS grain growth, especially beyond SLG substrates, is not well understood [22, 38, 44, 45]. In this work, we observed that Na from SLG significantly increases grain size notably on the surface compared to LNG, which is presumed to have minimal Na. Introducing additional Na via co-evaporation with 650 NaF yields optimal results and larger grains. However, increasing the NaF source temperature to 750 °C reduces grain size, which may result from pronounced vertical GGI grading as observed in **Fig. 1c**. Vertical GGI grading typically arises from limited Ga/In inter diffusion, leading to smaller grains, whereas minimal grading is associated with larger grain formation [42]. A similar trend of small grain formation with Na incorporation has been observed in CIGSe with steep Ga grading [29, 43] The correlation between grain growth and Na depth profile, shown in **Fig. 1b**, indicates that high Na concentration corresponds to regions with numerous grain boundaries, as seen in the SLG and 750 NaF absorbers. In contrast, 650 NaF, with fewer grain boundaries, shows lower Na concentration. This observation aligns with atomic probe tomography studies, which report an accumulation of Na at grain boundaries in CIGSe [46].

To investigate crystal quality, we analyzed the absorbers using X-ray diffraction (XRD). **Fig. 2b** presents the fine scan normalized θ-2θ XRD patterns for the 112 reflex, with the full-width at half maximum (β) values displayed. The wide scan XRD pattern (10-60°) is shown in **Fig. S5**. All peaks correspond to the tetragonal chalcopyrite structure of CIGS (ICDD:00-056-1309). The β value is lowest for 650 NaF, indicating superior crystallinity, which is consistent with SEM observations. This relation of β values to grain size is also noticeable for LNG, SLG, and 750 NaF absorbers: smaller grains in SEM correlate with larger β values. For 750 NaF, the peak becomes slightly asymmetric, and the 112 peak shifts to a higher angle, a trend also observed in the 700 NaF sample (Fig. S5). This shift is attributed to unit cell deformation (out-of-plane), potentially caused by strain release from the Mo surface, similar to observations in CIGSe on GaAs substrates with $Na_2Se$ PDT treatment [33]. In epitaxial CIGSe, a positive shift in the dominant 200 reflex after $Na_2Se$ treatment is related to out-of-plane unit cell parameter changes due to strain release. In our polycrystalline CIGS, the positive shift is attributed to release of strain caused by the Mo substrate with excess Na supply, which could explain the peeling effect observed with high (>750°C) NaF concentrations. To further elucidate this strain effect, we measured XRD before and after exfoliating a CIGS absorber (without NaF co-evaporation) from the Mo back contact, as shown in **Fig. S5**. After exfoliation, the Mo diffraction peak is absent, while the 112 CIGS peak shows a positive shift due to released strain. This effect is similar to that observed with the 750 NaF absorber, where excess Na accumulates on the Mo, releasing strain at the Mo/CIGS interface and impacting crystallization. A positive shift may also result from high Ga content due to lattice contraction from smaller Ga atoms partially occupying In lattice sites [7, 21, 47]. However, as Ga/In composition values are consistent on average across all absorbers, the positive shift is attributed primarily to strain effect.

We employed PL and CL techniques to evaluate the effect of Na on the optoelectronic properties of CIGS absorbers. These techniques are instrumental in screening absorber quality and understanding defect nature before proceeding with complete solar cell fabrication [7, 13, 48]. Absolute PL measurements help predict the maximum $V_{OC}$ and fill factor (FF) achievable in a completed solar cell [12, 13], while CL hyperspectral imaging can reveal the microscopic origins of recombination processes at the surface and across the absorber's depth [48, 49]. **Fig. 3a** presents the normalized PL spectra of the absorbers under 1 sun equivalent photon flux, while **Fig. S6** shows the absolute PL spectra. The plan-view mean CL spectra of the absorbers are shown in **Fig. 3b**. The $E_g$ derived from the PL maximum (denoted as $E_{g(PL)}$), which corresponds to band-to-band (BB) radiative emission, is approximately 1.59 eV for LNG, SLG and 650 NaF, and 1.61 eV for 750 NaF, as shown in **Fig. 3c**. Similarly, the CL maxima (denoted as $E_{g(CL)}$ corresponding to BB or near band edge emission (NBE)), are shown in **Fig. 3c**. Both PL and CL show nearly the same $E_g$ for all the absorbers, except for LNG. The differences in $E_g$, and defect emission, in LNG are attributed to inhomogeneities in the absorber. The absence of the dominant defect peak in CL is due to the significantly higher excitation intensity in CL compared to 1-sun PL. Higher excitation level makes the BB emission more prominent, as seen in the intensity-dependent PL spectra for LNG shown in **Fig. S6**, where BB emission dominates with increasing excitation intensity. It is important to note that excitation with 1-sun equivalent photon flux and understanding of the corresponding electronic transitions are crucial for evaluating solar cell performance. Typically, $E_{g(PL)}$ is expected to correlate with the minimum band gap (the notch) of the absorber's depth profile and can be described quantitatively with respect to GGI ratio [1]. The samples SLG and 650 NaF are expected to have similar or closely matched notch band gap, based on the GGI depth profile and ratios (Fig. 1a and 1c), which is further confirmed by the energy of the PL maxima (Fig. 3c). In contrast, the absorbers LNG and 750 NaF show the reverse trend in PL maximum energy than expected from the GGI profile. Here, LNG and 750 NaF show lower and higher $E_g$ despite having higher and lower GGI at notch, respectively. In the case of the LNG absorber this can be simply due to a shift of the PL maximum towards lower energies due to the high background from the defect luminescence and due to a shift of the emission energy due to tail states [13, 50]. The higher emission energy in the 750 NaF sample can indicate, that the main emission is not coming from the notch, due to low mobility, which prevents the carriers from reaching the notch, or it could just be due to inhomogeneities of the sample. To further confirm, we excited the 750 NaF absorber with a low energy (660 nm) laser, which can penetrates deeper into the absorber, and observed a ~50 meV blueshift in $E_{g(PL)}$ as shown in Fig. S6.

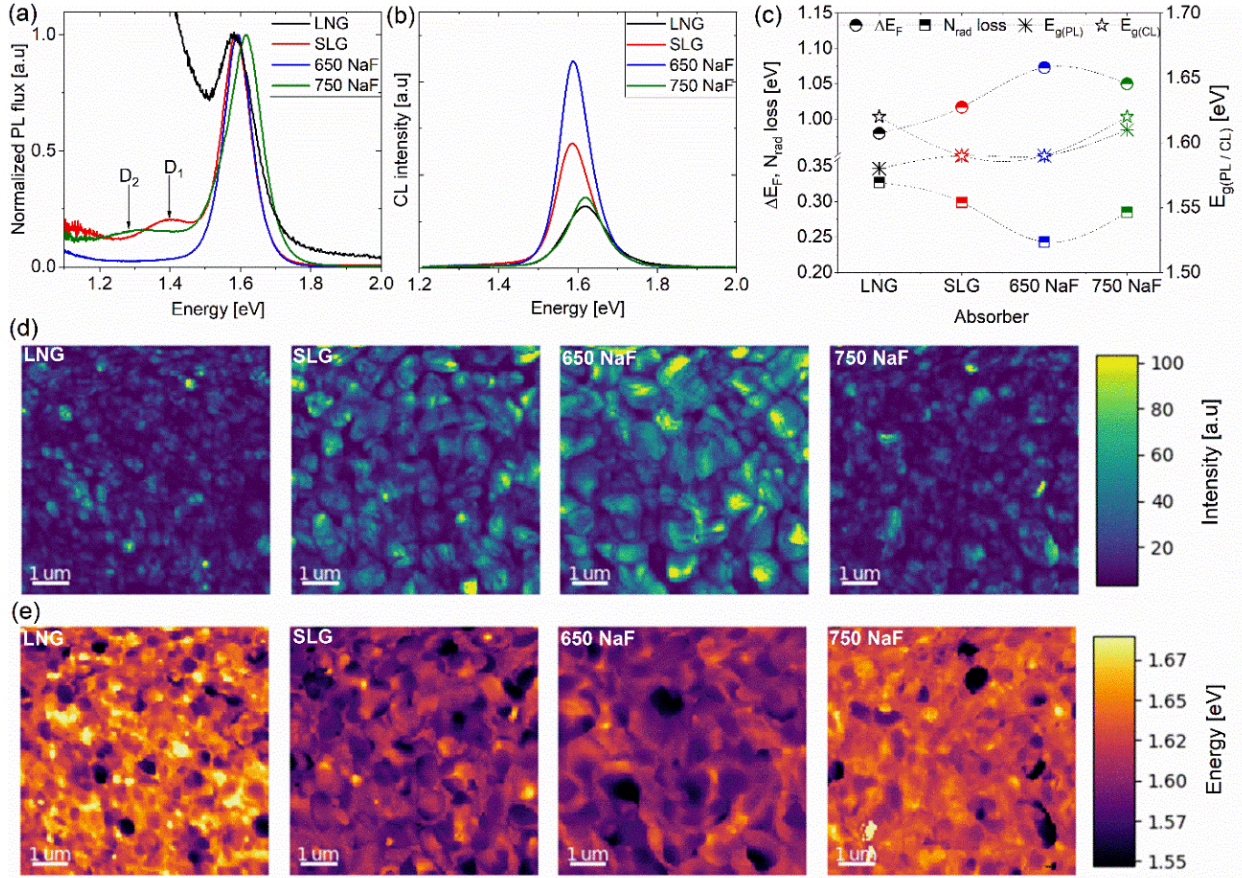

***Fig. 3:*** *Optoelectronic properties of CIGS with varying Na supply. (a) PL spectra, normalized to near band edge emission. (b) plan-view mean CL spectra. (c) optoelectronic parameters extracted from the 1 sun absolute PL measurements. The $E_{g(PL)}$ and $E_{g(CL)}$ for SLG and 650 NaF are overlapped on each other. (d) Panchromatic CL intensity maps and (e) NBE energy distribution maps of CIGS.*

LNG, with minimal Na, exhibits a broad deep defect ($D_2$) peak at lower energies (see log scale in Fig. S6), as defined in [7], indicating predominant recombination through defect states [47]. In contrast, SLG, with sufficient Na, shows dominant BB emission along with a weak and shallow defect ($D_1$) around 1.4 eV, suggesting that the Na supply during CIGS growth plays a critical role in reducing defects. The addition of 650 NaF further suppresses the $D_1$ defect, resulting in the highest PL quantum yield ($Y_{PL}$) among all absorbers. Absorbers with 600 and 700 NaF exhibit similar BB and $D_1$ peaks as SLG (**Fig. S6 and S7**). However, 750 NaF leads to a detrimental $D_2$ emission below 1.3 eV, adversely affecting photovoltaic performance, as also here observed by the lower PL quantum yield [7]. The mean CL spectra in **Fig. 3b** reflect NBE intensity variations among different absorbers roughly consistent with PL results (see Fig. S6). We used $Y_{PL}$ approach to determine the quasi-Fermi level splitting $\Delta E_F$ as described in Ref [12]. The $Y_{PL}$ was calculated using the relation, $Y_{PL}=\Phi_{PL}/\Phi_{laser}$, where $\Phi_{PL}$ represents an integrated flux of the BB emission peak between 1.45 eV and 2.0 eV, and $\Phi_{laser}$ represents an incident photon flux corresponding to the number of photons in a 1 sun spectrum above the $E_g$. This approach neglects reflection of the laser on the sample surface and thus underestimates the $Y_{PL}$ by about 20% [13]. The increase in $Y_{PL}$ with 650 NaF demonstrates a reduction in non-radiative losses. The $\Delta E_F$ is determined using the relationship of the Shockley-Queisser $V_{OC}$ limit and non-radiative recombination, as detailed in [51] and expressed in the following equation (1),

$$\Delta E_F = qV_{OC}^{SQ} + k_B T \times \ln Y_{PL} \qquad (1)$$

where $qV_{OC}^{SQ}$ is the SQ $V_{OC}$ limit, $k_B$ is the Boltzmann constant, and $T$ is the temperature. The $E_g$ for $qV_{OC}^{SQ}$ is taken from $E_{g(PL)}$, however, we note that this approach further underestimates the $\Delta E_F$ by typically 10 to 20meV [13]. The $E_{g(PL)}$, $\Delta E_F$ and non-radiative loss ($k_B T \times \ln Y_{PL}$) are shown in **Fig. 3c**. The $\Delta E_F$ for the 650 NaF absorber is 1.07 eV, which is 60 meV higher than the reference SLG absorber, suggesting a potential for achieving high $V_{OC}$ with an appropriate buffer layer [7, 10, 52]. We attribute the decrease in non-radiative losses ($N_{rad}$ loss) from 299 meV (SLG) to 243 meV (650 NaF) to the suppression of bulk defects, as supported by the PL spectra, a crucial effect for highly efficient solar cells [10]. Conversely, the $\Delta E_F$ for the LNG-only absorber is 0.98 eV, reflecting defect-mediated recombination losses, underscoring the importance of adequate Na supply during CIGS growth. However, excess 750 NaF leads to decrease in $\Delta E_F$ due to an increase in non-radiative loss.

**Fig. 3d** presents panchromatic plan view CL images of CIGS absorbers with different Na contents, LNG and 750 NaF show high-intensity emission from only a few grains, while 650 NaF exhibits a nearly uniform emission intensity across all grains with the overall highest intensity, consistent with PL results. CIGS on SLG shows uniform emission across grains but with lower total intensity. Dark regions in all maps indicate strongly inhibited radiative recombination in certain grains (notably high for LNG and 750 NaF) and at grain boundaries [7, 48, 53]. By fitting Gaussian functions to all peaks in each pixel of the maps we can extract the emission energy of the peaks and investigate their variation across the probed area, as presented in **Fig. 3e**. LNG is found to have a high NBE energy standard deviation of 27 meV, ranked the highest in all four samples. SLG and 650 NaF have a considerably lower standard deviation of the emission energy, at 19 meV and 18 meV, suggesting a much more uniform bandgap across the absorber surface. The NBE energy standard deviation for 750 NaF is 21 meV, which is much improved compared to LNG. Overall, these plan view CL results suggest that Na addition can enhance radiative recombination and improve compositional homogeneity.

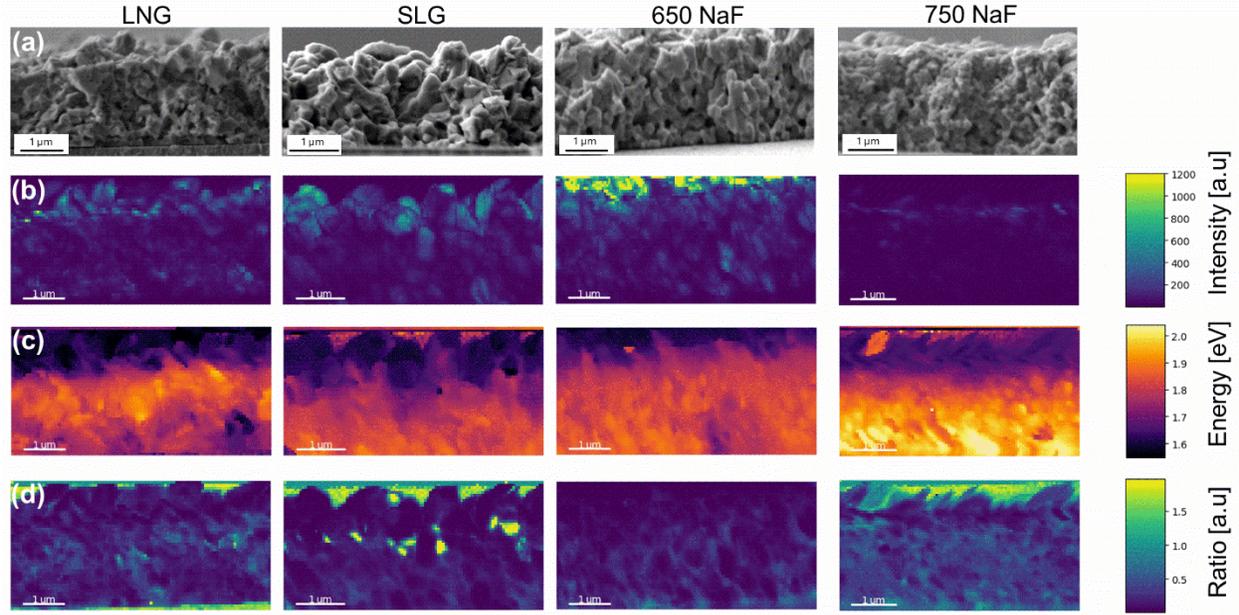

**Fig. 4:** *Cross-sectional hyperspectral CL analysis of CIGS with different Na supply. (a) Secondary electron images used for mapping. (b) Panchromatic intensity images. (c) Emission energy distribution across the absorber depth. (d) Defect density, extracted as the ratio of defect intensity peak over NBE intensity peak. The scale bar in all images is 1 µm.*

To gain a deeper understanding of the emission energy and defect density variations across the depth of the absorbers, we performed microscopic mapping of their cross-sections. **Fig. 4a** shows the cross-sectional SE images of the absorbers used for hyperspectral CL mapping. **Fig. 4b** provides panchromatic CL images showing the variation of the spectrally integrated intensity across the depth of the thin films. The mean CL spectra for cross-sectional CL are shown in **Fig. S8**. LNG and SLG show strong emission intensity at the surface and at approximately 0.5 µm sub-surface, a depth we associate roughly with the position of the notch, based on the GDOES data in **Fig.1c**. The CIGS absorber with optimal Na supply (650 NaF) exhibits strong emission from the surface to around 0.3 µm depth, gradually fading towards the back contact. In contrast, 750 NaF shows poor intensity across the depth except at the position of the notch, with negligible emission at the surface, reflecting the higher defect concentration. **Fig 4c** shows the distribution of NBE emission energy across the cross-sectional map. The emission energy map shows that the area with the lowest emission energy – corresponding to the notch – is located close to the surface, with an emission energy of ~1.6 eV for all the absorbers, consistent with $E_{g(PL)}$ and $E_{g(CL)}$. A clear variation in emission energy is visible towards the back of the absorbers. The emission energy as a function of depth, extracted from a line scan in **Fig. 4c**, is shown in **Fig. S9**. The trend of increasing emission energy towards the back aligns with the GGI profile observed in GDOES, as discussed in **Fig.1c**. However, the high surface GGI observed in LNG is not detected in the CL depth profile, likely due to inhomogeneity within the absorber.

To get insight into the defect distribution, we map the intensity of the defect emission relative to the NBE emission intensity (**Fig. 4d**). Note that the bright yellow areas at the top of the maps are due to places, where the electron beam hits the empty space above the surface of the film not the actual cross section. These regions are artefacts which do not give information about the defect distribution. The LNG and even more the SLG film show spots with very high defect emission. In particular, in the SLG film they seem to be located near the steepest gradient in the emission energy (see Fig. 4c). All films besides the 650 NaF film show some defect emission near the surface, albeit reduced compared to the bulk. The 650 NaF film shows very low defect emission near the surface. It is likely that a low near surface defect density contributes to a reduction in space-charge region (SCR) recombination [53] and thus a lower diode factor in comparison with 750 NaF (see Table S2). Furthermore, the absence of surface defects helps to improve the translation of $\Delta E_F$ into actual $V_{OC}$ [52].

Additionally, we extracted spectra along the lines marked in **Fig. S10** to further understand emission energy variations across the depth, as shown in **Fig. 5a.** It is worth noting that the line scan spectra integrated all the CL signals surrounded marker within 2.5 µm in width. The line spectra of LNG reveal jumps in the emission energy (best visible as peak broadening and splitting in the extracted spectra) near the notch position (marked with a red box). Jumps in the band gap, as opposed to a smooth gradient, were shown to be detrimental to absorber quality [54]. Additionally, LNG shows a band gap gradient with lower bandgaps towards the back contact, which will certainly increase back contact recombination [17, 55, 56] . The lower bandgap near the back contact in the LNG sample is visible in many areas in the map in **Fig. 4c**. Such a jump in the NBE emission energy near the notch position also can be observed in SLG and 650 NaF sample. No clear reverse bandgap gradient in the middle or lower part of the absorber has been found in the line spectra of SLG and 650 NaF. One area shown in the map in Fig. 4c exhibited a reverse gradient, which may be caused by local inhomogeneity. 750 NaF has the smoothest emission energy transition around the notch position with only subtle NBE peak change, but it, overall, has the largest energy change from the frontside to the backside (agreeing with the GDOES measurements). Similar to the observation the top view CL maps, a moderate amount of Na addition can improve overall compositional uniformity.

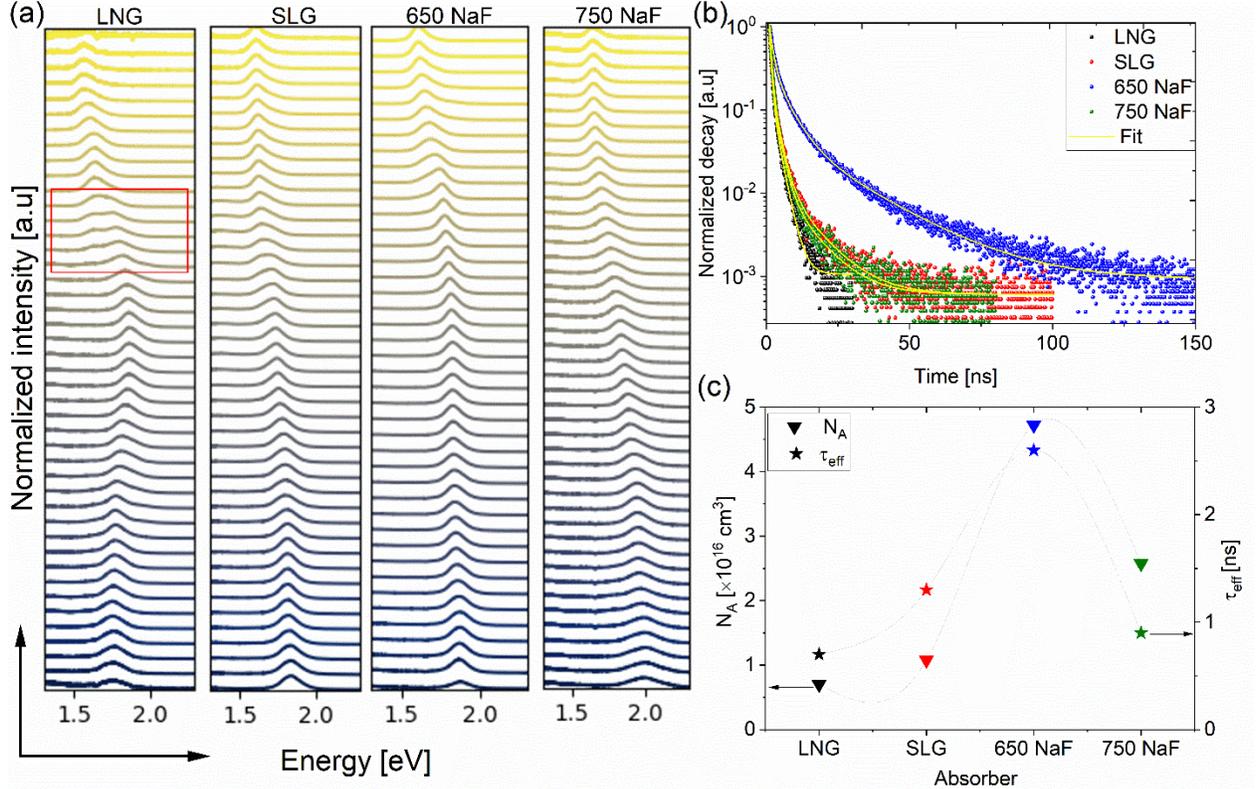

**Fig. 5:** *Cross-sectional CL line scan of CIGS with varying Na supply, showing variation in spectra across the depth. (b) TRPL transients and (c) calculated doping density alongside effective lifetime of CIGS with different Na supply.*

In **Fig. 4c** we can also see pronounced variations in emission energy across the film, corresponding to regions of higher and lower gallium content. In **Fig. S11**, we show another set of line spectra, averaging over a much narrower region (0.5 μm). These spectra show distinct jumps in emission energy with depth, and double emission peaks in places where regions of more than one composition are being sampled. This also illustrates the simultaneous existence of regions of higher and lower gallium content, consistent with previous observations in [54]. This indicates that further process optimization is needed to grow phase pure CIGS.

Increased quasi-Fermi level splitting can be due to longer minority carrier lifetimes and due to higher doping [12]. Time resolved photoluminescence (TRPL) spectroscopy gives an indication on changes in minority carrier lifetime. **Fig. 5b** presents the TRPL transients for the absorbers, measured at the lowest excitation intensity possible to ensure that we measure minority carrier lifetimes. The decays are not exponential, therefore we fit them with a tri-exponential fit (a bi-exponential fit for LNG). The fitted values for the TRPL transients are summarized in **Table S1**. The long-lifetime component ($\tau_3$) of the TRPL transients, which accounts for both surface and bulk recombination, is notably high for the 650 NaF-based CIGS, reaching up to 20.6 ns. This feature indicates a longer carrier lifetime and potentially improved charge extraction efficiency. The effective carrier lifetime, representing the weighted average of time constants, is calculated using the relation described in Equation (2) [57],

$$\tau_{eff} = \frac{\sum A_i \tau_i}{\sum A_i} \tag{2}$$

where, the $A_i$s are the amplitudes of the different lifetimes $\tau_i$, $i$=1,2,3. The $\tau_{eff}$ is plotted in **Fig. 5c**. $\tau_{eff}$ is considerably smaller than long decay time ($\tau_2$ or $\tau_3$, see Table S1) for all samples, because it also contains the fast decay processes at the beginning of the decay. However, they both show the same trends. The long decay time for LNG is 2.4 ns which increases to 9.7 ns for SLG. The optimal 650 NaF further increases the $\tau_3$ to 20.6 ns while the excess 750 NaF declines the $\tau_3$ to 9.0 ns. The short decay time ($\tau_1$) observed in all the absorbers is attributed to the charge carrier separation at the surface or trapping [57, 58]. Charge carrier dynamics in CIGS have not been extensively explored in literature, with only a few recent reports providing insights [7, 16]. For instance, previous work from our laboratory demonstrated a long decay time of 3.2 ns, which improved to 4.5 ns with the addition of a buffer layer [7]. The notably long decay ($\tau_3$), without buffer layer, observed in this study for CIGS with 650 NaF is attributed to reduced non-radiative losses, a lower defect density across the absorber's depth, and an increased grain size. To gain a deeper understanding of the different contributions to the increase in $\Delta E_F$, we calculated the hole doping density ($N_A$) using the relationship between electron Fermi level and lifetime on the one side and hole Fermi level and doping density on the other side (see e.g. [16]), as described by equation (3),

$$N_A = p_0 = \frac{d.N_C N_v}{G.\tau_3} exp\left(\frac{\Delta E_F - E_g}{k_B T}\right) \qquad (3)$$

where, $N_c$ and $N_v$ are the effective density of the states of the conduction and valance band, where we use the values for CuInS$_2$ [15] as it is closeset available, and Ga content is consistant on average across all the absorbers. $G$ is the generation flux during steady state $\Delta E_F$ measurement and $d$ is the thickness of the film (~2.8 μm from cross sectional SEM, see Fig. 4). The calculated doping density is presented in **Fig. 5c**. The $N_A$ for the LNG absorber is $7.0\times10^{15}$ cm$^{-3}$, which increases to $1.08\times10^{16}$ cm$^{-3}$ in the SLG grown absorber. The optimal Na supply from 650 NaF further enhances the $N_A$ to $4.7\times10^{16}$ cm$^{-3}$. However, with higher 750 NaF supply, the $N_A$ decreases to $2.5\times10^{16}$ cm$^{-3}$. The observed trend in doping concentration is consistent with findings from various studies on chalcopyrite absorbers grown with an alkali barrier and controlled Na supply [23, 25, 38]. Analysis of the best-case scenario for SLG and 650 NaF absorbers reveals that a ~0.6 fold increase in $N_A$ and a ~2 ns increase in long decay time contributes to a ~10 meV improvement in $\Delta E_F$, considering the same $E_g$.

One of the reasons for the variation in hole doping density could be attributed to the Na out-diffusion process during CIGS growth, as theoretically observed in selenide chalcopyrite [32]. The available Na supply during the CIGS growth promotes the formation of Na$_{Cu}$ impurities, due to its low formation energy (0.55 eV) relative to Cu in the chalcopyrite structure. However, Na$_{Cu}$ is thermodynamically unstable, and Na migrates out of the grains with a migration energy of 0.36 eV at room temperature [59], leaving behind Cu vacancies, which results in an increased $N_A$. This phenomenon, however, is not fully understood in the context of high Na concentrations within the grains, as we observe a decrease in $N_A$ with increasing Na supply. The decrease in $N_A$ with excess NaF could be linked to the poor quality of the absorber, as discussed earlier. It is important to note that the optimal NaF concentration has a very narrow window. For instance, with 600 and 700 NaF, the Na supply is either below or above the optimal range, necessitating precise optimization depending on experimental conditions. This narrow window is due to the significant variation in NaF evaporation rate with every 20 °C change in source temperature.

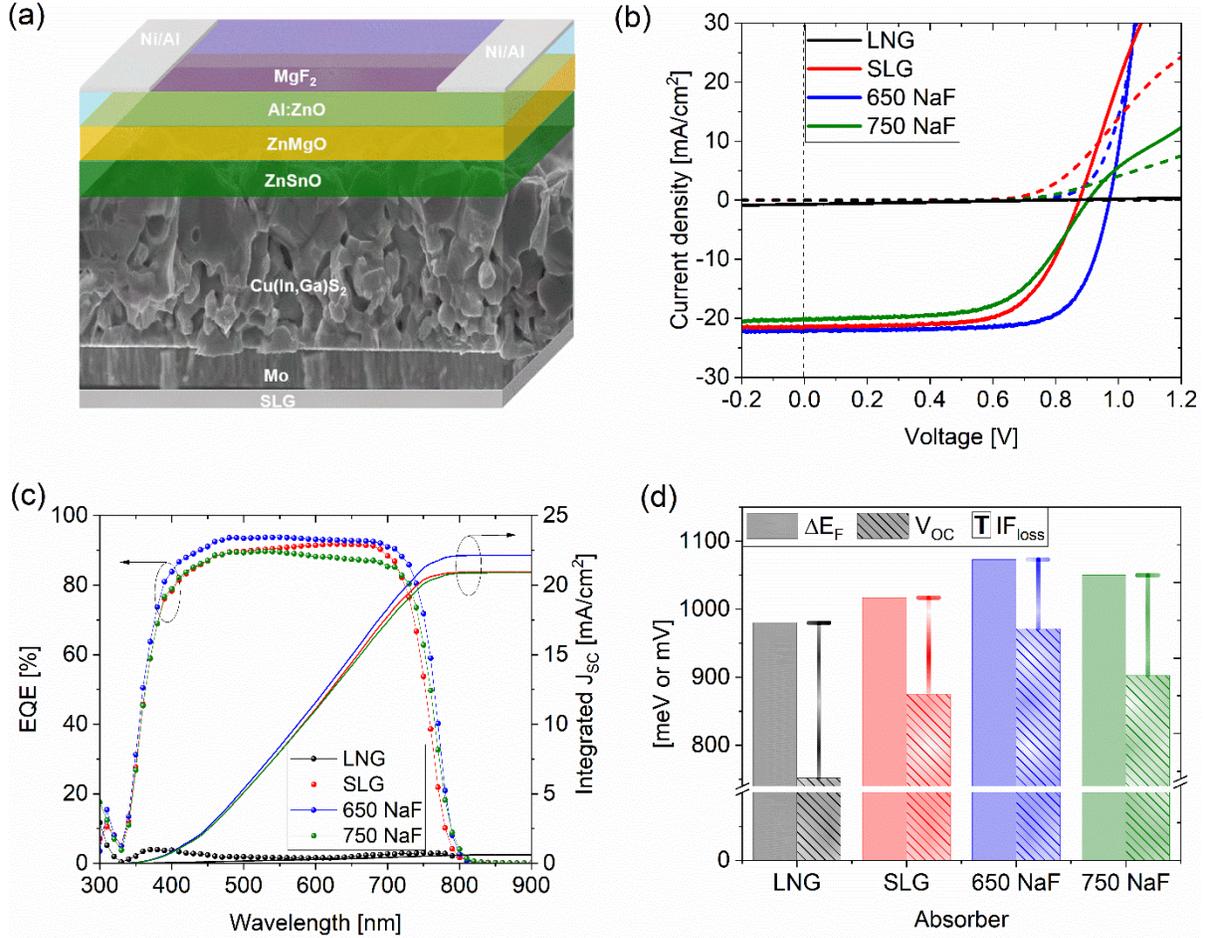

**Fig. 6:** *Photovoltaic performance of the CIGS solar cell with varying Na supply. (a) Schematic of the solar cell structure with cross-sectional SEM image of 650 NaF absorber. (b) Active area J-V curves under dark and illuminated conditions. (c) EQE spectra alongside integrated $J_{SC}$. (d) interfacial loss determined from the difference between $ΔE_F$ and $V_{OC}$.*

Finally, we fabricated the solar cells incorporating ~50 nm thick ZnSnO buffer layer, deposited using atomic layer deposition, in combination with a sputtered $(ZnMg)O_x$ i-layer, to avoid barriers for the forward current and fill factor losses [19]. Thus, the device configuration is given by Mo/CIGS/ZnSnO/(ZnMg)$O_x$/Al:ZnO/MgF$_2$ as illustrated in Fig. 6a. The top three layers (from top to bottom) function as an antireflection coating, a transparent front contact, and an intrinsic layer, respectively. Additionally, some unintentional variations in the buffer layer composition, particularly in the [Sn]/([Sn]+[Zn]) ratio, were observed due to gas flow interruptions on the surface of the NaF co-evaporated CIGS. The [Sn]/([Sn]+[Zn]) ratios for devices based on LNG and SLG substrates were 0.19, measured on reference SLG pieces within the deposition chamber, while for NaF 650

**Table 1:** *Photovoltaic parameters $V_{OC}$, $J_{SC}$, FF and PCE of the best CIGS solar cells with varied Na supply during the growth. The PCE values were calculated based on both $J_{SC(EQE)}$ and $J_{SC(JV)}$ are provided.*

| Absorber | Voc (mV) | Jsc(EQE) (mA/cm²) | Jsc(J-V) (mA/cm²) | FF (%) | PCE (EQE) (%) | PCE (J-V) (%) |
|---|---|---|---|---|---|---|
| LNG | 752 | 0.6 | 0.4 | 27.2 | 0.1 | 0.1 |
| SLG | 875 | 20.9 | 21.3 | 66.2 | 12.1 | 12.3 |
| 650 NaF | 971 | 22.1 | 22.0 | 73.1 | 15.7 | 15.7 |
| 750 NaF | 903 | 20.9 | 20.2 | 59.9 | 11.3 | 10.9 |

and NaF 750, the ratios were 0.15. **Fig. 6b** presents the active area current density-voltage (J-V) curves under dark and illuminated conditions, and **Fig. 6c** displays the external quantum efficiency (EQE) alongside the integrated short-circuit current density ($J_{SC}$) of the best performing solar cells. The corresponding photovoltaic parameters ($J_{SC}$, $V_{OC}$, FF, and PCE) are summarized in **Table 1**. The PCE was calculated based on the integrated $J_{SC}$ extracted from the EQE spectrum; this is the preferred approach, owing to the spectral mismatch of the halogen lamp used in the solar simulator. Furthermore, the dark I-V curves were fitted using the standard one-diode and two-diode models and the corresponding series resistance, shunt resistance, diode factor and dark current values are summarized in Table S2. The 650 NaF based solar cell shows reduced series resistance and diode factor in comparison with 750 NaF solar cell. The interfacial loss ($IF_{loss}$), determined from the difference between the $\Delta E_F$ and $V_{OC}$, is shown in **Fig. 6d**. The solar cell fabricated on the LNG substrate exhibited poor performance, as anticipated from the 1 sun PL spectrum and defect density map from CL mapping. Despite all parameters being suboptimal for the LNG device, a reasonable $V_{OC}$ of 752 meV was obtained. The $J_{SC}$ however is almost non-existent. The best solar cell based on SLG delivered a PCE of 12.1% with a $V_{OC}$ of 875 meV. Despite the high $\Delta E_F$, defects at or near the surface lead to increased non-radiative recombination and a gradient in the electron quasi-Fermi level, resulting in a lower $V_{OC}$ [52]. We note that, the performance of the SLG based device could be less optimal compared to other NaF co-evaporated devices, based on the different [Sn]/([Sn]+[Zn]) ratios, which likely cause differences in the band alignment. As anticipated from the optoelectronic and structural characterization results, the 650 NaF based solar cell emerged as the champion, delivering a high $V_{OC}$ of 971 meV with an FF of 73.1%, resulting in an active area

PCE of 15.7%. Furthermore, in the optimal 650 NaF device the interface deficit was significantly decreased to 102 meV. Our previous work has shown, that further buffer optimization can reduce this deficit further [7]. However, increasing the Na supply to 750 NaF proved detrimental to solar cell performance, despite yielding a better $V_{OC}$ than the SLG-based device. The increased $V_{OC}$ deficit in the 750 NaF device, compared to the 650 NaF device, is possibly due to surface defect density. The drop in PCE is mainly attributed to the poor FF and $J_{SC}$, as widely observed in CIGSe with excess Na [35, 40]. The kink observed in the J-V curve is indicative of a potential barrier at the interface, which likely results from non-optimized band alignment, indicating once more the need for further buffer optimization [19, 60]. This barrier can also cause voltage dependent carrier collection and maybe the explanation of the reduced EQE and $J_{SC}$ [19]. In the case of the 750 NaF based solar cell, the kink is more pronounced, suggesting that the interface barrier may be more significant, thereby contributing to reduced fill factor and overall efficiency. The average values of the photovoltaic parameters follow the same trends as those of the best cells. **Fig. S12** shows the statistical distribution of photovoltaic parameters from 8 devices (4 for LNG). For 650 NaF, the solar cells exhibit a narrow distribution of photovoltaic parameters, indicating better homogeneity and absorber quality.

The $E_g$ values extracted from the inflection point of the EQE spectra for all absorbers are presented in **Fig. S13.** The EQE spectrum for LNG was poor and is therefore excluded from further discussion. Compared to the CIGS with 650 NaF the SLG sample shows lower EQE across the wavelength (λ) range of 400-720 nm, indicating a lower collection function near the front surface. This could be due to low doping [19] and a large SCR, which would shift the p-n point far into the absorber and make holes minority carriers near the front interface [55]. The decreased EQE for 750 NaF, compared to 650NaF, may be due to voltage dependent carrier collection due to a barrier. **Fig. S13** shows the voltage dependent EQE spectra of the solar cells. The carrier collection remains unaffected under negative bias for all the solar cells. However, a significant reduction is observed under positive bias for the SLG and 750 NaF based solar cells, while 650 NaF solar cells shows relatively minimal reduction. The unaffected shape of the EQE spectra indicates that the carrier collection loss is independent of wavelength. However, electrical $J_{SC}$ losses are more pronounced in SLG and 750 NaF solar cells. In summary, optimizing NaF supply during CIGS growth is crucial for suppressing surface and bulk defects, which is essential for achieving highly efficient solar cells with optimally band-aligned buffer layers.

**Conclusions**

In this study, we have elucidated the critical role of additional Na supply in optimizing the performance of wide bandgap CIGS photovoltaic absorbers. By combining Na from the glass substrate with NaF co-evaporation, we demonstrated that optimal Na supply enhances Ga/In interdiffusion, reduces lateral bandgap fluctuations, and promotes the formation of larger grains with superior crystallinity. 1 sun calibrated PL measurement reveal that optimal Na leads to a significant reduction in non-radiative losses, resulting in a high $\Delta E_F$ of 1.07 eV and a long carrier decay time of 20.6 ns. Cross-sectional CL hyperspectral mappings further indicate that reducing defect density at the surface and SCR is crucial for effectively translating $\Delta E_F$ into a high $V_{OC}$. The champion wide bandgap solar cell fabricated with 650 NaF CIGS achieved an active area PCE of 15.7% and a $V_{OC}$ of 971 meV, underscoring the importance of precise Na management for achieving high performance CIGS solar cells.

**Experimental**

The CIGS absorbers were grown using a well-established 3-stage process previously described for both sulfide and selenide chalcopyrite [7, 56, 61]. We used Molybdenum (Mo) coated soda lime glass substrates (SLG) and low Na glass (LNG) substrate. The actual substrate temperatures ($T_S$) were ~ 470 ºC during the first stage and ~600 ºC during the second and third stages, with a ramp rate of 20 ºC/min and substrate rotation at 4 rpm. The actual substrate temperature was estimated based on the substrate heater calibration using pyrometer. The schematic representation of the 3-stage process is provided in **Fig. S1**. Before evaporation, the Mo surface was cleaned at $T_S$ of 450 ºC for 15 minutes followed by exposure to sulfur vapor at $T_S$ of 400 ºC for 15 minutes, resulting likely in the formation of a $MoS_2$ layer, although its thickness remains undetermined. In the first stage, calibrated fluxes of gallium (Ga) and indium (In) were co-evaporated under a sulfur partial pressure ranging between $2\times10^{-5}$ mbar and $8\times10^{-5}$ mbar. In the second stage, calibrated fluxes of copper (Cu) and sodium fluoride (NaF) were co-evaporated. NaF in all the absorbers was co-evaporated for 45 minutes regardless of Cu evaporation time (usually 50 minutes to 60 minutes). Once a Cu-rich stoichiometry was formed (~7.5 % excess, monitored by an increase in heating power and pyrometer readout), Ga and In were co-evaporated again (and no Cu) in the third stage to form a Cu-poor stoichiometry, with the Ga flux being lower than in the first stage [56]. Various batches of CIGS were grown by varying the NaF source temperature between 600 ºC, 650 ºC, 700 ºC, and 750 ºC [29, 36]. All the NaF co-evaporated CIGS absorbers were deposited on SLG substrates. The ZnSnOx buffer layer was deposited in an F120 Microchemistry ALD reactor at 120 °C using a metal-organic precursor/$N_2$-purge/$H_2O$/$N_2$-purge cycle with the pulse and purge times of 0.4/0.8/0.4/0.8 s respectively. The metal-organic precursor tetrakisdimethylamino-Sn (TDMASn) was heated to 40 °C in a bubbler source using $N_2$ carrier gas, whereas the other metal-organic precursor diethyl-zinc (DEZ) and $H_2O$ were effused into the reactor due to their high vapour pressures at room temperature. The ZnSnOx process used a five cycle supercycle, where the metal-organic precursor for each cycle was altered in the following sequence: DEZ, DEZ, TDMASn, DEZ, and TDMASn. This supercycle scheme resulted in a [Sn]/([Sn]+[Zn]) ratio of 0.18 for the $ZnSnO_x$ film, except for the process run where the gas flow was disrupted which led to a ratio of 0.15. The supercycle was repeated 300 times using a total of 1500 cycles, which resulted in a $ZnSnO_x$ film thickness of about 50-60 nm. This was followed by an Al:MgZnO window layer and an Al:ZnO transparent front contact, both deposited by magnetron sputtering. The device was completed with an evaporated Ni/Al grid electrode, topped with an ~100 nm thick $MgF_2$ antireflection coating. Individual cells are defined by mechanical scribing with an active area of ~0.42 $cm^2$.

**Characterizations**

The compositions were analyzed by energy-dispersive X-ray spectroscopy (EDX) at operating voltages of 7 kV and 20 kV. The compositional depth profiles were recorded by a glow discharge optical emission spectroscopy (GDOES). An argon plasma at a pressure of 450 Pa erodes the sample and generates a light emission of the sputtered atoms. The specific emission lines of the singles elements are optical diffracted and detected by several photomultipliers. For the quantification of the CIGS matrix elements, the reference measurement by EDS was used. The sodium concentration is determined by a relative sensibility factor (RSF), which was calculated from implanted reference samples.

The cross-sectional morphology of the absorbers was examined using scanning electron microscopy (SEM). Structural characterization was conducted through X-ray diffraction (XRD) utilizing Cu-Kα radiation and the data was plotted without applying instrumental resolution correction.

The quasi-fermi level splitting ($\Delta E_F$) measurements on the bare absorbers were performed by a home-built absolute PL set-up equipped with a 405 nm laser excitation source at room temperature using a CCD Si detector [12]. The experimental setup is first calibrated, and the resulting spectra are corrected spectrally and in intensity, as described in ref [54]. The incident photon flux has been set to match the 1 sun photon flux that an absorber of $E_g$ 1.6 eV would receive under a AM 1.5 illumination. The PL quantum yield ($Y_{PL}$) method was used to determine the $\Delta E_F$ using the PL maximum as the band gap [12, 13] and explained in the PL section.

The CL measurements were performed using an Attolight Allain 4027 Chronos dedicated CL-SEM at room temperature. All CL maps were acquired with 5 kV beam acceleration voltage, 1.25 nA measurement current, and 50 μm aperture. The exposure time per pixel for all maps was 500 ms. The interaction volume under the current setting estimated by Monte Carlo Casino has an extent of about 150 nm. The CL data analysis was carried out using Hyperspy and Lumispy [62, 63]. For the extraction of NBE and defect peak features, such as peak energy, intensity and full-width-half-maximum, two Gaussian peaks were applied during fitting.

The TRPL measurements were performed using a time-correlated single photon counting (TCSPC) system with a 638 nm pulsed laser operating at a repetition rate of 10 MHz. The band edge emission was detected with a bandwidth of 46 nm.

The solar cells were characterized using a class AAA solar simulator calibrated with a reference standard Si solar cell (RQN3154). Current-voltage (I-V) measurements were performed at room temperature in forward direction using a IV source measure unit, with a scan speed of 50 mV/s. The active solar cell area, approximately 0.42 cm², was determined using a Leica microscope and ImageJ software. The active area is given by the outer circumference of the scribed cell minus the area of the metal grids. The entire active area was illuminated with 1 Sun (100 mW/cm²) during the measurements without using any mask. The external quantum efficiency (EQE) of the solar cells was measured using chopped illumination from a halogen-xenon lamp (Xenon short arc lamp, Ushio UXL-302-0) and a lock-in amplifier to detect the photocurrent.

**Acknowledgement**

This work was funded by Luxembourgish Fond National de la Recherche (FNR) for REACH (Project no: INTER/UKRI/20/15050982) and European Union within the SITA project (no. 101075626). The Cambridge team would like to acknowledge funding from the EPSRC under EP/V029231/1 and EP/R025193/1. For the purpose of open access, the author has applied a Creative Commons Attribution 4.0 International (CC BY 4.0) license to any Author Accepted manuscript version arising from this submission. Funded by the European Union. Views and opinions expressed are however those of the author(s) only and do not necessarily reflect those of the European Union or CINEA. Neither the European Union nor the granting authority can be held responsible for them. We thank Thomas Schuler for the technical support and Dr. Damilola Adeleye for the discussion on the absorber growth.

# Sodium induced beneficial effects in wide bandgap Cu(In,Ga)S$_2$ solar cell with 15.7% efficiency

Arivazhagan Valluvar Oli[1], Kulwinder Kaur[1] Michele Melchiorre[1], Aubin Jean-Claude Mireille Prot[1], Sevan Gharabeiki[1], Yucheng Hu[2], Gunnar Kusch[2], Adam Hultqvist[3], Tobias Törndahl[3], Wolfram Hempel[4], Wolfram Witte[4] Rachel A. Oliver[2] and Susanne Siebentritt[1]

[1]Laboratory for Photovoltaics, Department of Physics and Materials Science Research Unit, University of Luxembourg, 41 rue du Brill, L-4422 Belvaux, Luxembourg
[2]Department of Materials Science and Metallurgy, University of Cambridge, 27 Charles Babbage Road, Cambridge CB3 0FS, UK
[3]Ångström Solar Center, Division of Solar Cell Technology, Department of Materials Science and Engineering, Uppsala University, 75121 Uppsala, Sweden.
[4]Zentrum für Sonnenenergie- und Wasserstoff-Forschung Baden-Württemberg (ZSW), Meitnerstraße 1, 70565 Stuttgart, Germany

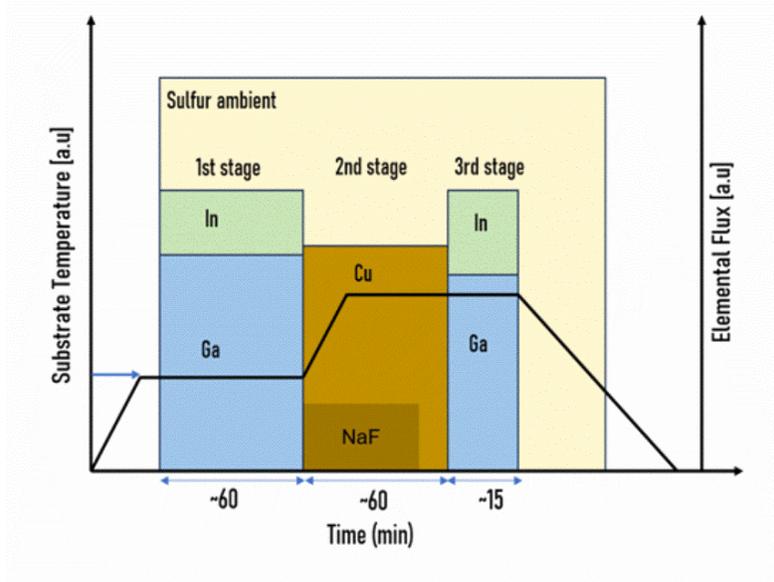

**Fig. S1:** Deposition profile of CIGS with NaF co-evaporation. The bars in the schematic are the corresponding elemental fluxes and the back thick line is the substrate temperature. The Mo coated substrates were annealed in the beginning of the process without Sulfur ambient to clean the surface.

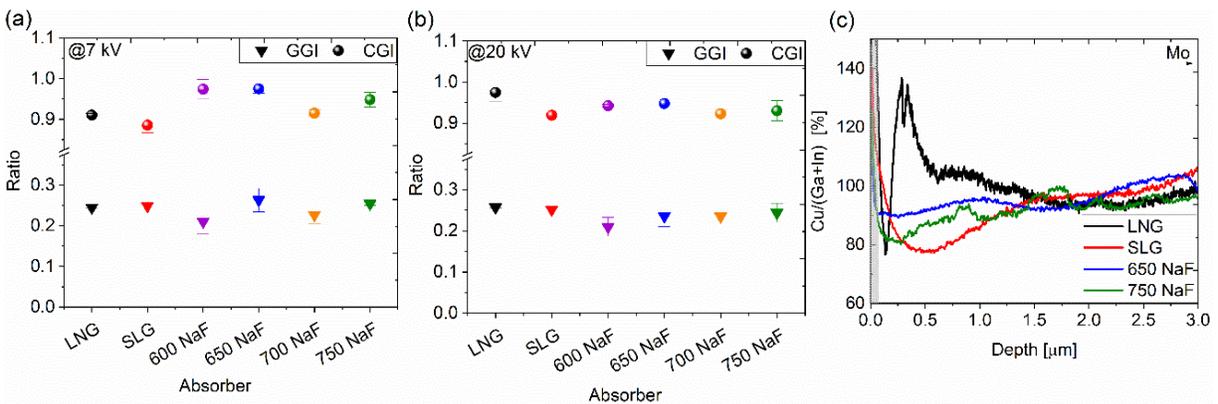

**Fig. S2:** GGI and CGI ratios of CIGS measured at the (a) surface (7 kV) and (b) bulk (20 kV). (c) CGI profile of the absorbers measured from GDOES. Related to main text **Fig. 1**.

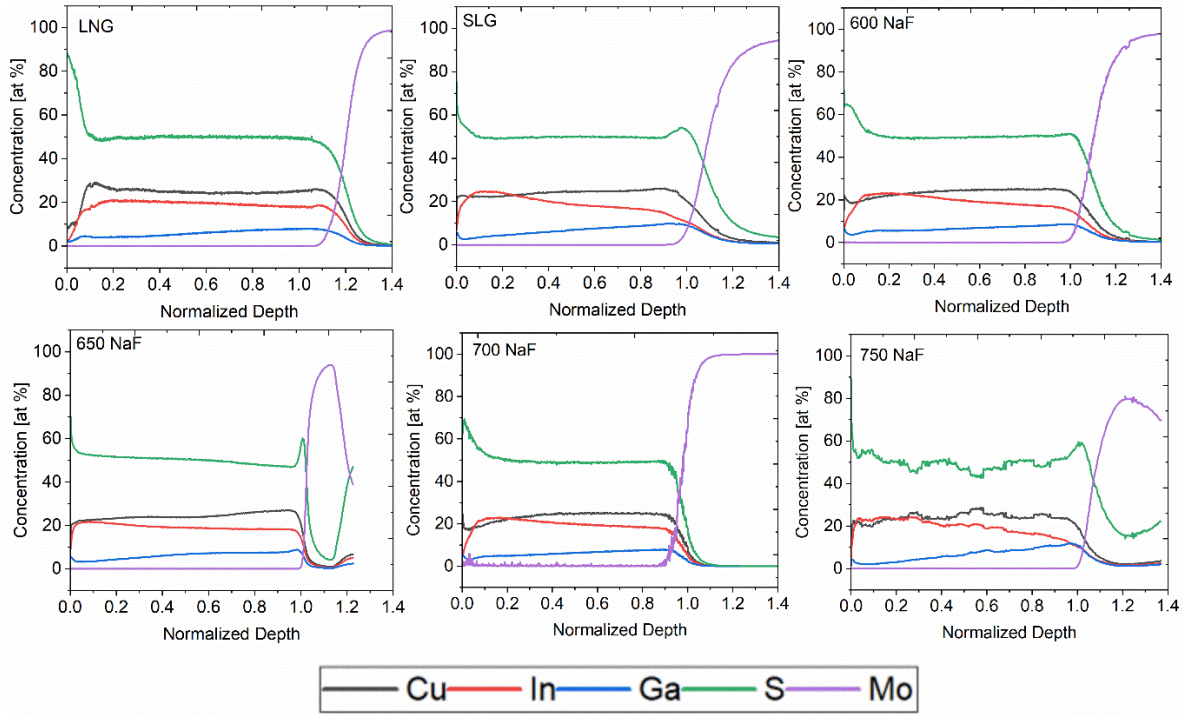

**Fig. S3:** GDOES depth profiles of various elements in CIGS with different Na supplies during growth. Related to main text **Fig. 1**.

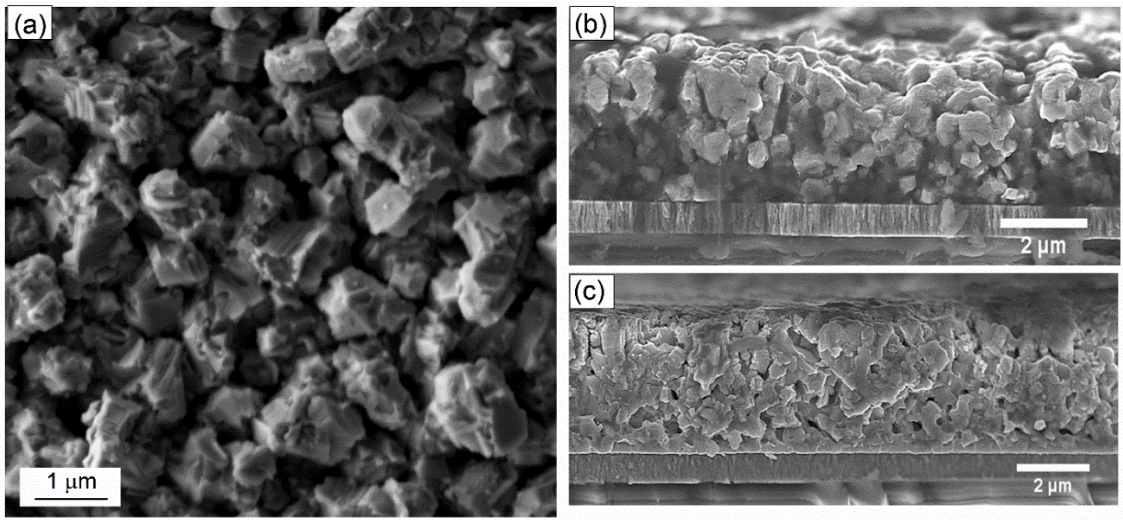

**Fig. S4:** (a) Top view SEM image of 750°C NaF CIGS. Cross-sectional SEM images of CIGS absorbers with (b) 600°C and (c) 700°C NaF. Related to main text **Fig. 2a.**

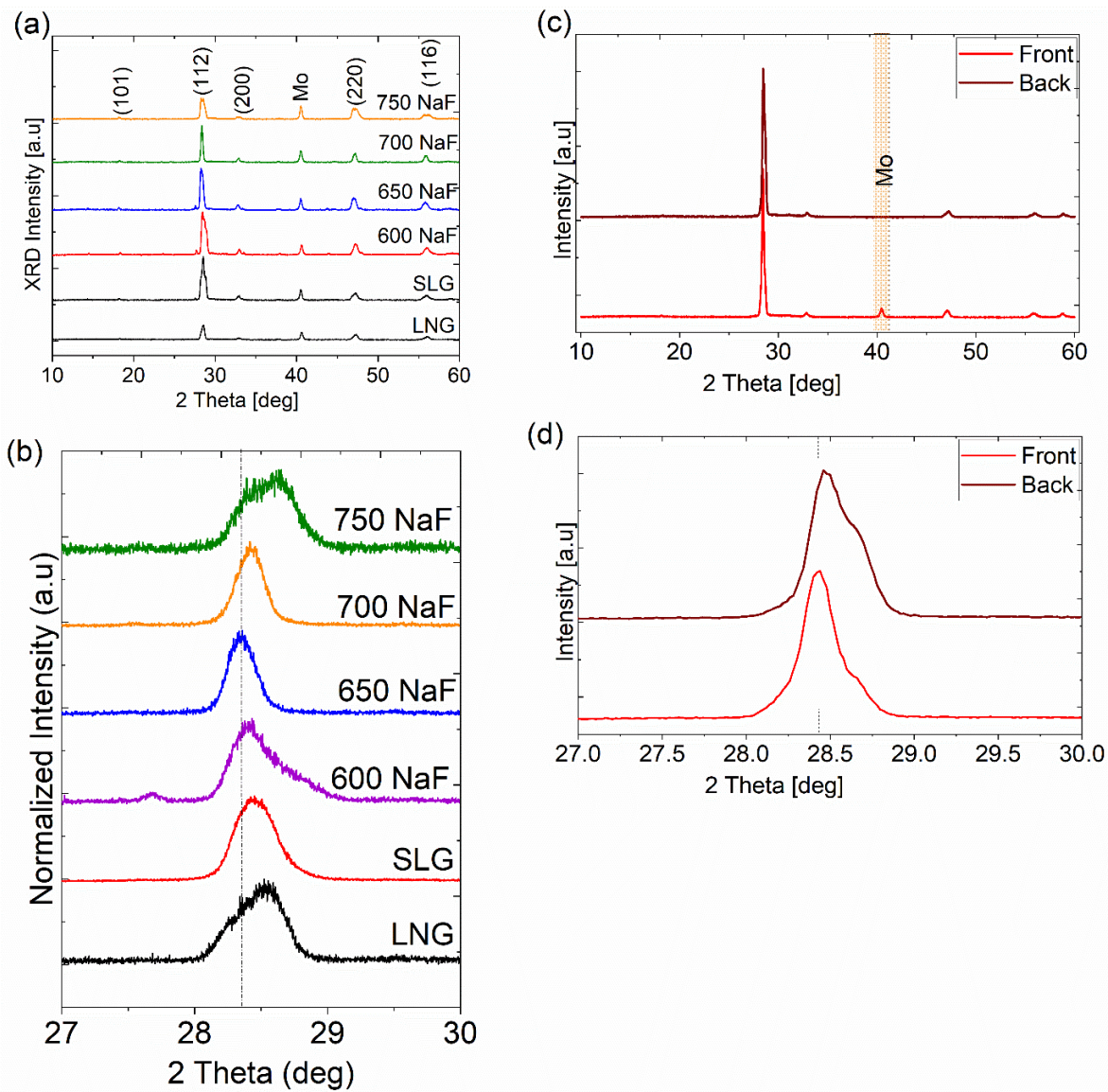

**Fig. S5:** (a) Wide-scan and (b) fine-scan X-ray diffractorgrams of CIGS with different Na supplies. (c) XRD patterns of CIGS (without NaF co-evaporation) before and after exfoliation from the Mo substrate. (d) Enlarged view of the dominant 112 reflex, showing a positive shift due to release of strain caused by the Mo substrate. The CIGS exfoliation process is described elsewhere [1, 2]. Related to main text **Fig. 2b.**

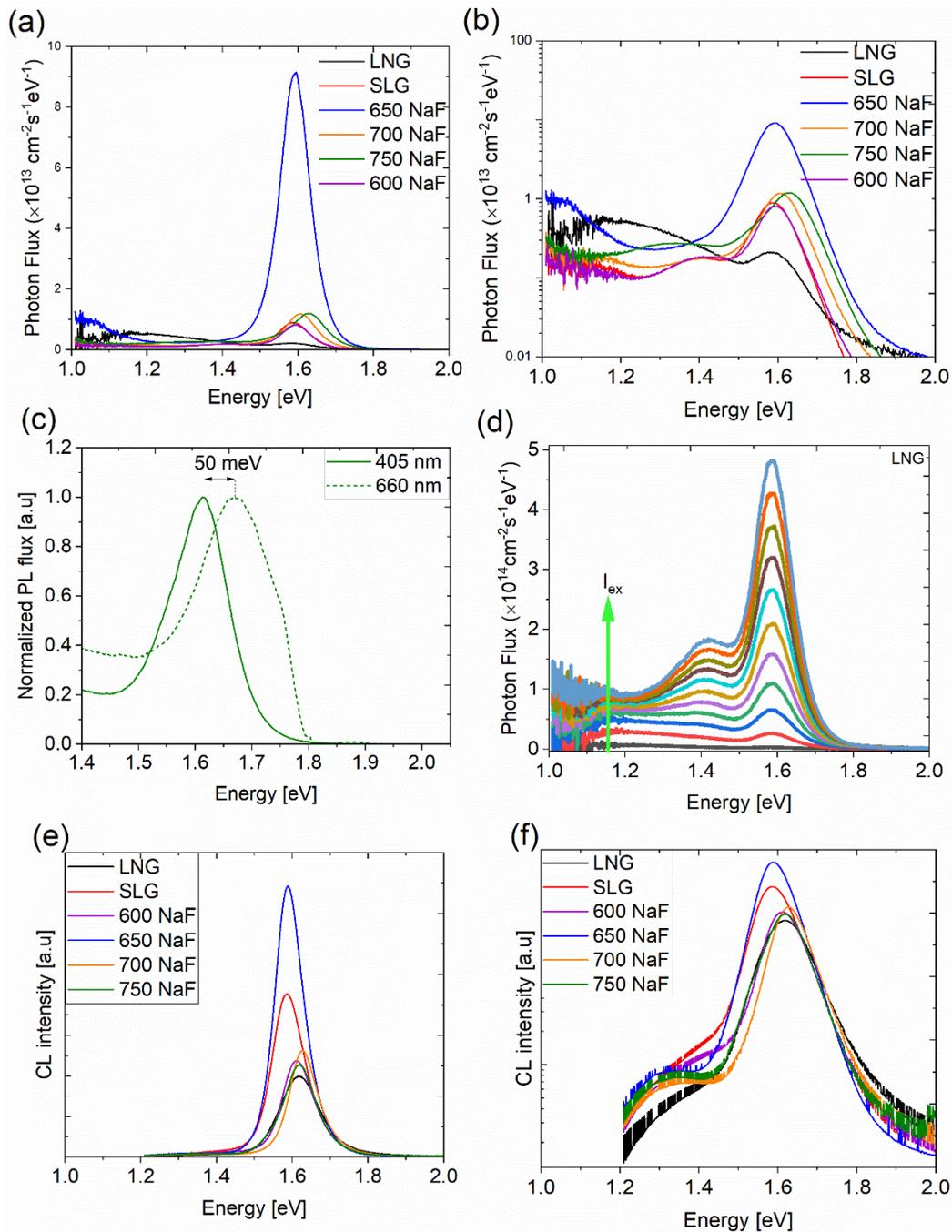

**Fig. S6:** (a) 1 Sun excited absolute PL spectra of CIGS with varying Na supplies and (b) the spectra in log scale to highlight low-intensity emissions. (c) PL spectra of 750 NaF absorber excited with 405 and 660nm laser, both with 25 mW power (d) Intensity-dependent PL spectra of LNG. CL spectra in (e) linear and (f) log scale. Related to main text **Fig. 3**.

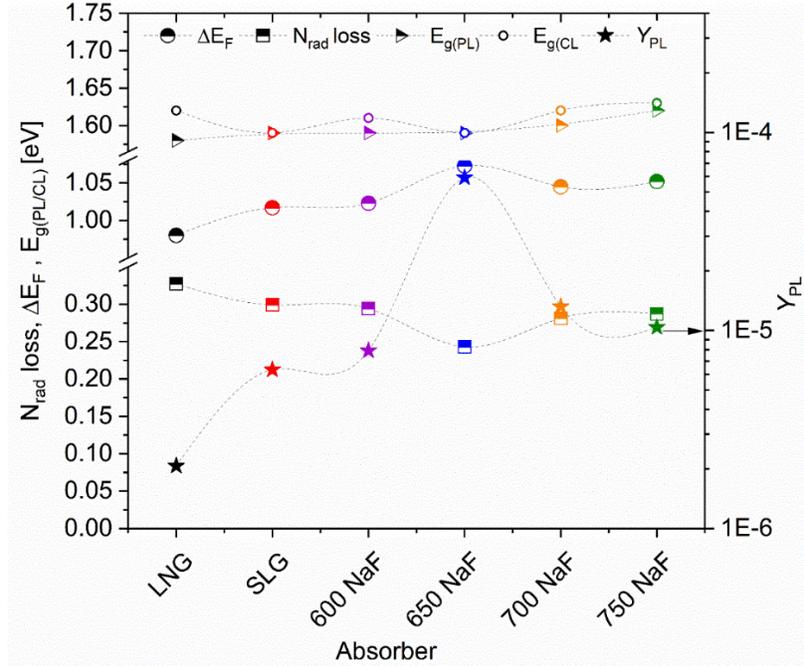

**Fig. S7:** Optoelectronic parameters derived from the PL and CL measurements.

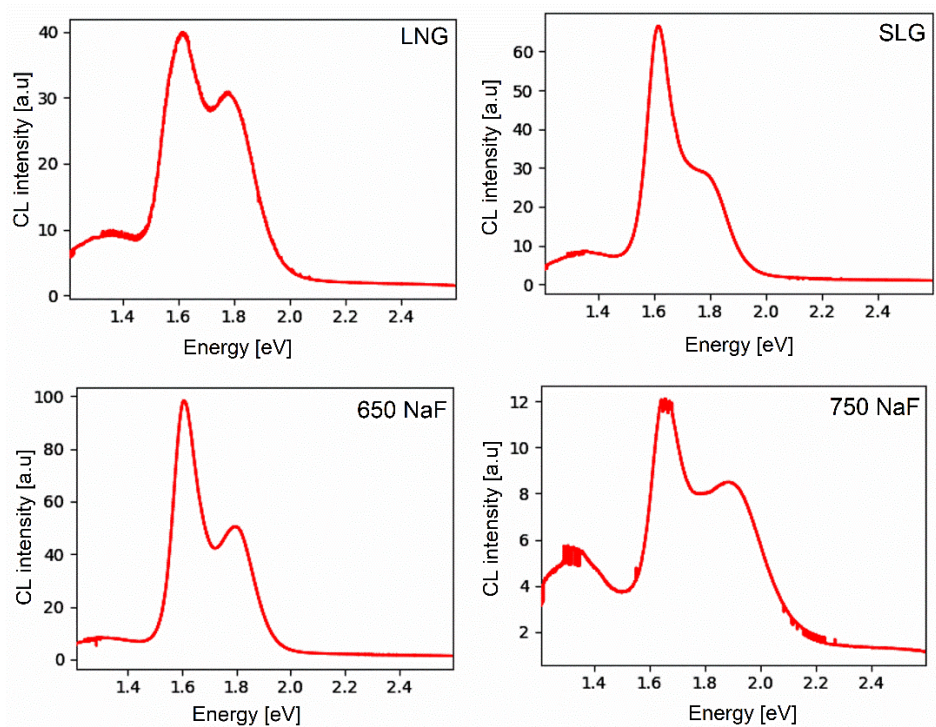

**Fig. S8:** Cross-sectional mean CL spectrum for LNG, SLG, 650 NaF and 750 NaF absorbers. Related to main text **Fig. 4.**

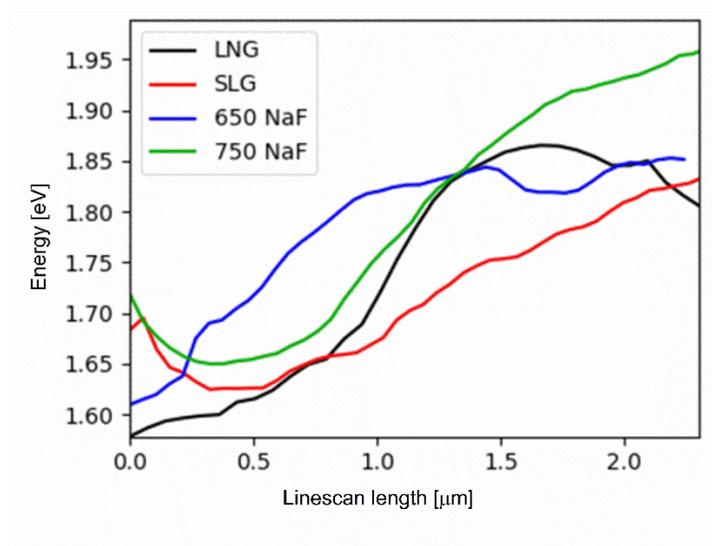

**Fig. S9:** Line scans extracted from the NBE energy map shown in Fig. 4c. The positions of line scan are as same as the positions in Fig. S9. The extracted profiles highlight the graded bandgap feature of the absorber. For 650°C NaF, no energy change is observed after the notch, which extends to the surface. The large surface ratio for SLG and 750 NaF is due to artifacts from the empty area above the surface. Related to main text **Fig. 4**.

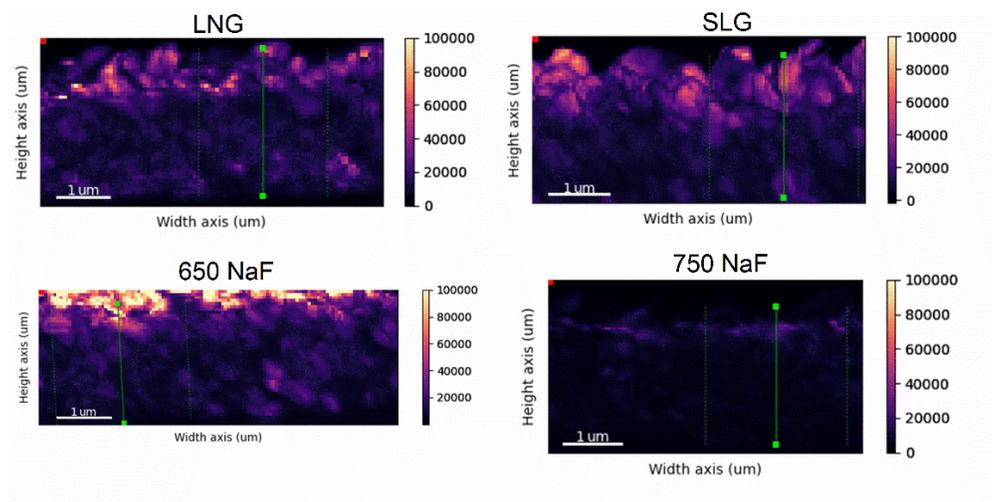

**Fig. S10:** Cross sectional panchromatic CL map with line scan markers. The width of line scan is 2.5 μm. Related to main text **Fig. 5a**.

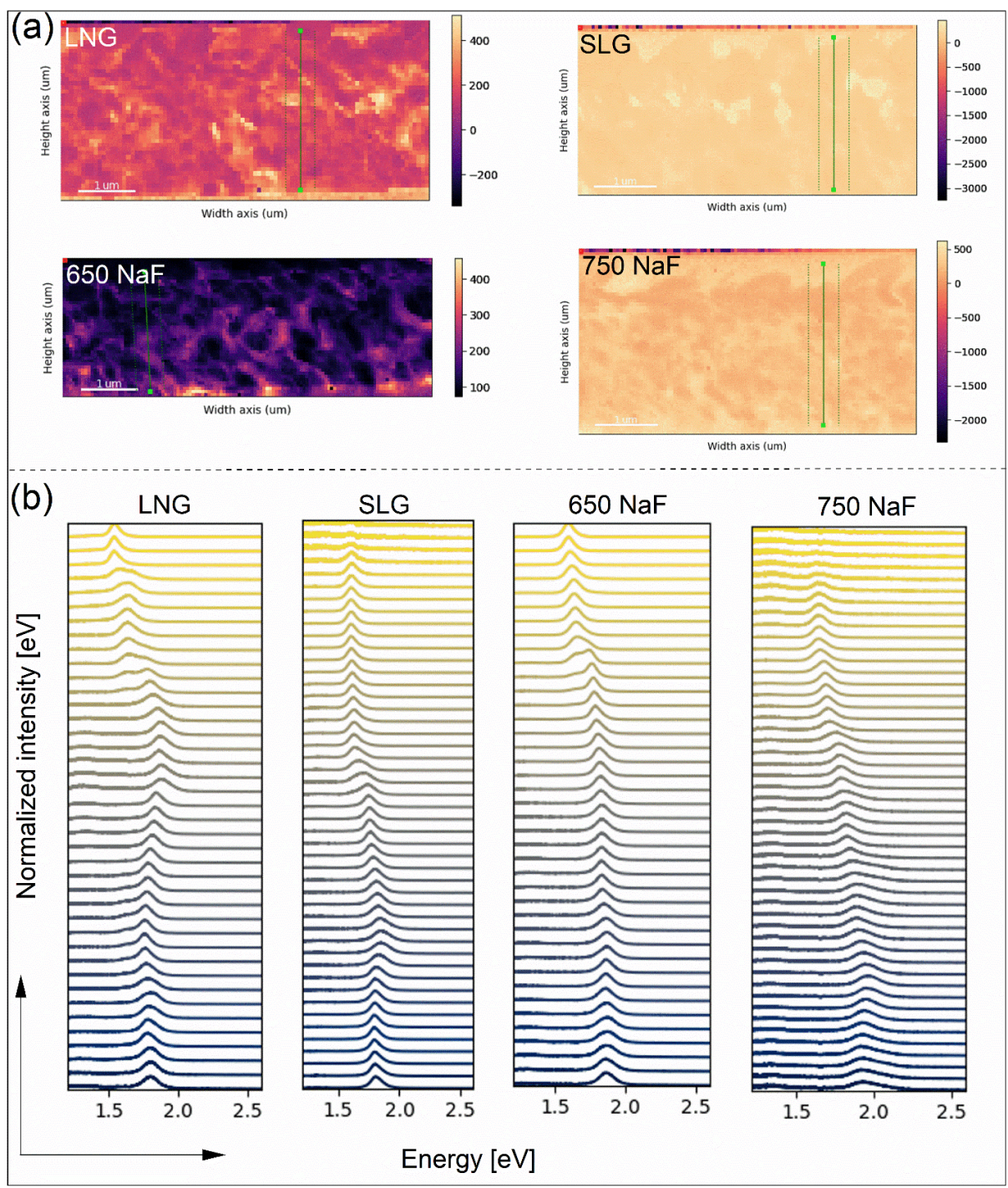

**Fig. S11: (a)** Cross sectional panchromatic CL map with narrow line scan markers. The width of line scan is 0.5 µm. (b) corresponding CL line scan spectra across the depth. Related to main text **Fig. 5a**.

TRPL curves were fitted using tri-exponential equation,

$$y = y_0 + A_1 \cdot \exp(-(x-x_0)/t_1) + A_2 \cdot \exp(-(x-x_0)/t_2) + A_3 \cdot \exp(-(x-x_0)/t_3)$$

For LNG, bi-exponential fit was used.

**Table S1:** TRPL fitted parameters. "A" represents the pre-factors, and "τ" represents different lifetimes. Related to main text **Fig. 6a**.

| Absorbers | LNG | SLG | 650 NaF | 750 NaF |
|---|---|---|---|---|
| A1 | 4.2 | 0.25 | 0.35 | 1.05 |
| A2 | 0.24 | 0.58 | 1.46 | 0.23 |
| A3 | - | 0.017 | 0.056 | 0.01 |
| $\tau_1$ [ns] | 0.62 | 2.18 | 5.06 | 0.61 |
| $\tau_2$ [ns] | 2.42 | 0.58 | 1.23 | 2.07 |
| $\tau_3$ [ns] | - | 9.72 | 20.68 | 9.08 |
| $\tau_{eff}$ [ns] | **0.7** | **1.3** | **2.6** | **0.9** |

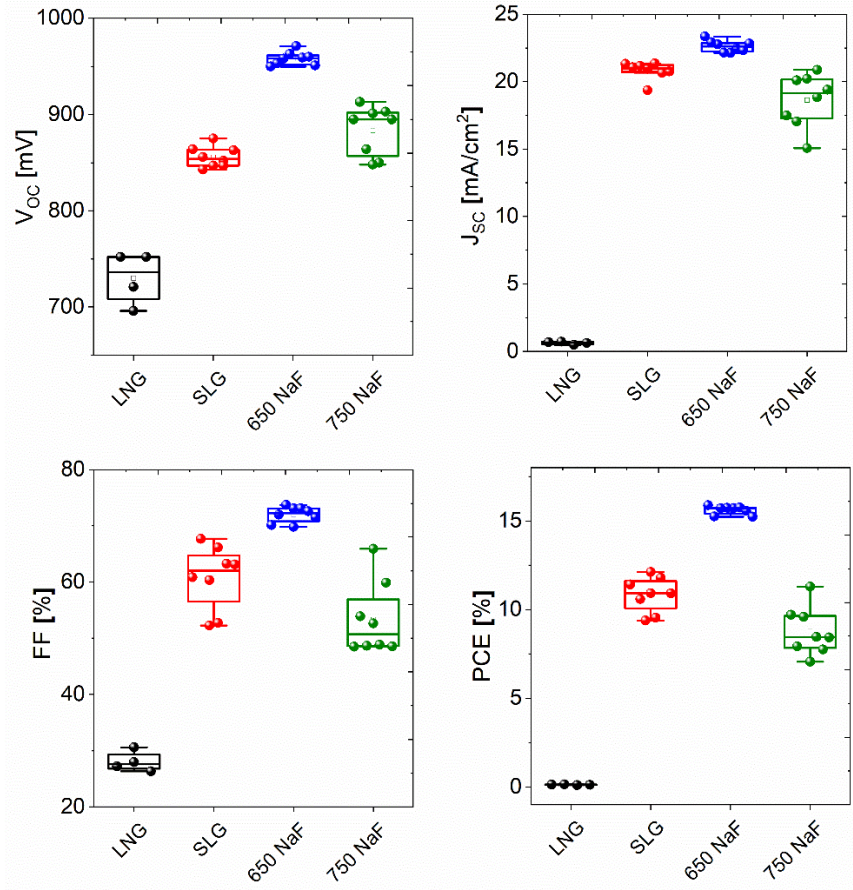

**Fig. S12**: Statistical distribution of the photovoltaic parameters of the solar cells. Related to **Fig. 7** in main text.

**Table S2:** *Fitted dark J-V parameters for series resistance ($R_s$) Shunt resistant ($R_{sh}$) diode factor (A) and dark current ($J_0$). Note that the buffer composition for LNG and SLG is different than 650 and 750 NaF.*

| Cell | $R_s$ ($\Omega.cm^2$) | $R_{sh}$ ($\Omega.cm^2$) | A | $J_0$ (A/cm$^2$) |
|---|---|---|---|---|
| LNG | 538 | Nano ohm.cm$^2$ | 11.8 | $2.55\times10^{-6}$ |
| SLG | 6.2 | 96833 | 2.75 | $3.26\times10^{-8}$ |
| 650 NaF | 0.2 | 12261 | 2.8 | $1.13\times10^{-8}$ |
| 750 NaF | 23.8 | 26147 | 2.9 | $2.39\times10^{-8}$ |

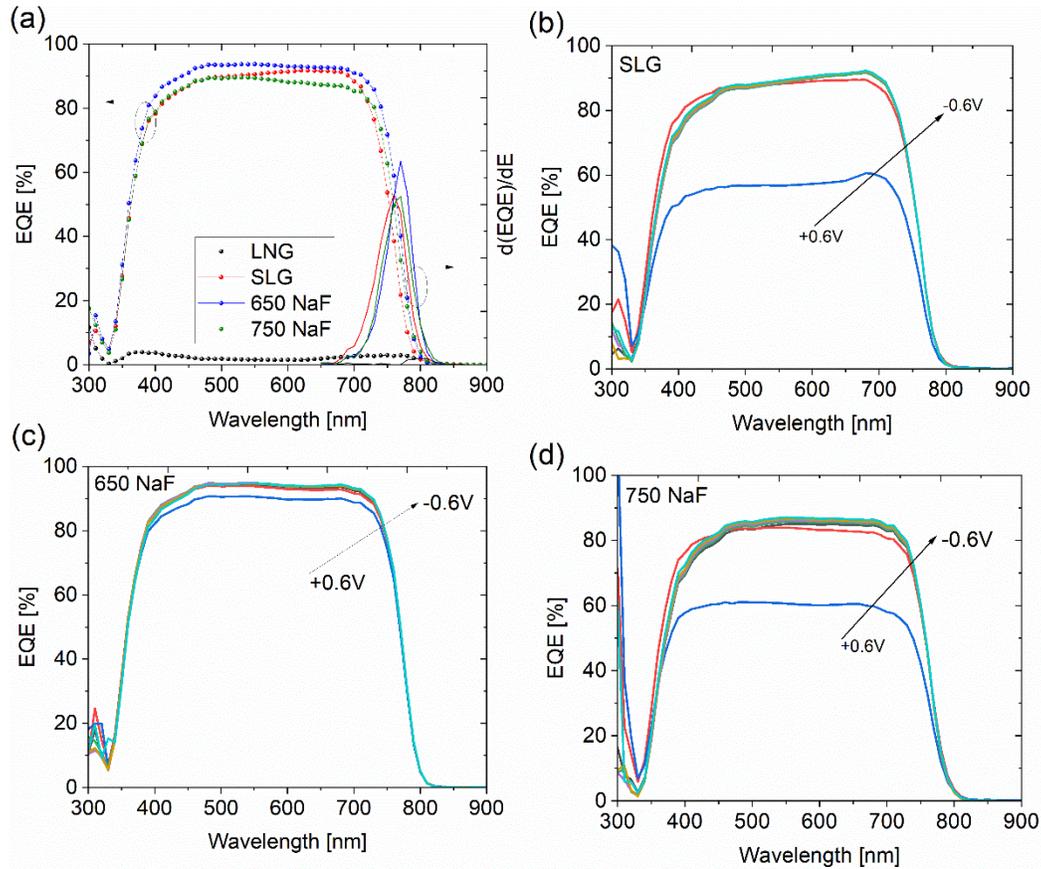

**Fig. S13:** (a) EQE spectra of CIGS with varying Na supplies, alongside $E_g$ extracted from the inflection point. Voltage dependent EQE spectra of (b) SLG, (c) 650 NaF and (d) 750 NaF absorber. Related to main text **Fig. 7b**.